\documentclass[twocolumn,trackchanges]{aastex701}
\shorttitle{Dusty galaxies in A2744: 3.3$\mu \rm m$ PAH bright galaxies selected from F430M emitters}
\shortauthors{Cheng, Wang, Liang et al.}

\begin{document}

\title{\large Probing Obscured Star Formation in Galaxy Clusters Using JWST Medium Band Images: 3.3$\mu\rm m$ PAH Emitter Sample in Abell 2744}

\author[0000-0003-0202-0534]{Cheng Cheng}
\affiliation{Chinese Academy of Sciences South America Center for Astronomy, National Astronomical Observatories, CAS, Beijing 100101, People’s Republic of China, chengcheng@nao.cas.cn, lpr@nao.cas.cn}
\affiliation{CAS Key Laboratory of Optical Astronomy, National Astronomical Observatories, Chinese Academy of Sciences, Beijing 100101, People’s Republic of China}
\email[show]{chengcheng@nao.cas.cn}

\author[0000-0002-9373-3865]{Xin Wang}
\affiliation{School of Astronomy and Space Science, University of Chinese Academy of Sciences (UCAS), Beijing 100049, China, xwang@ucas.ac.cn}
\affiliation{National Astronomical Observatories, Chinese Academy of Sciences, Beijing 100101, China}
\affiliation{Institute for Frontiers in Astronomy and Astrophysics, Beijing Normal University, Beijing 102206, China}
\email[show]{xwang@ucas.ac.cn}

\author[0000-0001-9143-3781]{Piaoran Liang}
\affiliation{Chinese Academy of Sciences South America Center for Astronomy, National Astronomical Observatories, CAS, Beijing 100101, People’s Republic of China, chengcheng@nao.cas.cn, lpr@nao.cas.cn}
\affiliation{CAS Key Laboratory of Optical Astronomy, National Astronomical Observatories, Chinese Academy of Sciences, Beijing 100101, People’s Republic of China}
\email[show]{lpr@nao.cas.cn}

\author[0000-0002-4622-6617]{Fengwu Sun}
\affiliation{Center for Astrophysics $|$ Harvard \& Smithsonian, 60 Garden St., Cambridge, MA 02138, USA}
\email{fengwu.sun@cfa.harvard.edu}

\author[0009-0008-9801-2224]{Edo Ibar}
\affiliation{Instituto de F\'isica y Astronom\'ia, Universidad de Valpara\'iso, Avda. Gran Breta\~na 1111, Valpara\'iso, Chile}
\affiliation{Millennium Nucleus for Galaxies (MINGAL)}
\email{eduardo.ibar@uv.cl}

\author[0000-0002-0245-6365]{Malte Brinch}
\affiliation{Instituto de F\'isica y Astronom\'ia, Universidad de Valpara\'iso, Avda. Gran Breta\~na 1111, Valpara\'iso, Chile}
\email{malte.brinch@uv.cl}

\author[0000-0001-7592-7714]{Haojing Yan} 
\affiliation{Department of Physics and Astronomy, University of Missouri, Columbia, MO 65211}
\email{yanhaojing@gmail.com}

\author[0000-0001-6511-8745]{Jia-Sheng Huang}
\affiliation{Chinese Academy of Sciences South America Center for Astronomy, National Astronomical Observatories, CAS, Beijing 100101, People’s Republic of China, chengcheng@nao.cas.cn, lpr@nao.cas.cn}
\affiliation{CAS Key Laboratory of Optical Astronomy, National Astronomical Observatories, Chinese Academy of Sciences, Beijing 100101, People’s Republic of China}
\affiliation{Harvard-Smithsonian Center for Astrophysics, 60 Garden Street, Cambridge, MA 02138, USA}
\email{jhuang@nao.cas.cn}

\author[0000-0001-9328-4302]{Jun Li}
\affiliation{Center for Astrophysics, Guangzhou University, Guangzhou 510006, People's Republic of China}
\email{lijunphy@gmail.com}

\author[0000-0002-8136-8127]{Juan Molina}
\affiliation{Instituto de F\'isica y Astronom\'ia, Universidad de Valpara\'iso, Avda. Gran Breta\~na 1111, Valpara\'iso, Chile}
\affiliation{Millennium Nucleus for Galaxies (MINGAL)}
\email{juan.molinato@uv.cl}

\begin{abstract}
Star-forming galaxies in galaxy clusters play a crucial role in understanding the advanced stages of galaxy evolution within dense environments. We present a sample of 3.3$\mu$m PAH-bright galaxies in the Abell 2744 (A2744) galaxy cluster. Using F430M medium band images, we select PAH emitters in the galaxy cluster, which capture the 3.3$\mu$m PAH emission at the redshift of A2744. Our multi-wavelength study demonstrates consistent star formation rates (SFRs) derived from PAH emission and SED fitting, indicating the 3.3 $\mu$m PAH flux estimated from medium band image alone can reveal the entirety of star formation, immune to dust obscuration. We find that the PAH emitters are located in relatively low mass surface density regions of A2744, with SFRs aligning with the field star-forming main sequence at $z=0.3$. The PAH emission morphologies show more asymmetry than that of the F444W image when asymmetry index $> 0.4$. With these results, we suggest that these star-forming galaxies in A2744 are in the stage of falling into the cluster from the field, and have not been quenched yet. We further explore a potential link between these galaxies and cosmic filaments being accreted onto the cluster, which may channel gas inflows to fuel star formation. JWST medium-band imaging provides a powerful new tool for identifying heavily dust-obscured star-forming populations. Future HI and low-J CO observations should be prioritized to resolve the cold gas kinematics and star formation processes in these systems, which would directly test the role of environmental stripping versus filamentary gas supply. 

\end{abstract}



\section{Introduction} \label{sec:intro}
Galaxy clusters are the most massive gravitationally bound systems, with a halo mass of about $M_{\rm halo}\sim 10^{14} M_\odot$, and are usually considered the most extreme overdense environments for galaxy evolution. According to the environmental quenching scenario, the massive halo will accrete and shock-heat the inflowing cold gas, quenching star formation in the cluster's central region \citep[e.g.,][]{1977MNRAS.179..541R, 2010ApJ...721..193P}. The morphology-density relation of galaxies in clusters shows a high fraction of elliptical galaxies with low star-formation rates (SFRs) in the cluster center \citep{1980ApJ...236..351D, 1997ApJ...490..577D}. The ram pressure exerted by the intracluster medium (ICM) and the tidal disruption from nearby massive galaxies will strip gas from the galaxies, further assisting the environmental quenching process, making galaxy clusters very harsh environments for star formation.

On the other hand, how star formation persists in galaxy clusters, and whether the star formation activity is similar to that in field galaxies, remains debated. If star formation is a local process that only depends on the molecular gas, then star formation activity in cluster galaxies would be similar to that in field galaxies. Meanwhile, the morphology density relation, and other results also show the existence of environmental quenching \citep{1978ApJ...219...18B, 1995A&A...298..361J, 1996MNRAS.279....1B, 1996A&A...314..379P, 2018NatAs...2..695M, 2025arXiv250201724D}. Therefore, comparing star-forming galaxies in clusters with those in the field will enhance our understanding of star formation in different environments, as well as the stage of galaxy evolution in clusters. 

Moreover, galaxy clusters are accreting cold gas and field galaxies through filaments \citep{2007A&A...470..425B, 2021MNRAS.503.2065K, 2021ApJ...923..235C, 2020NatAs...4..900Q, 2020MNRAS.494.5473K}. And for A2744, previous studies have demonstrated the filamentary structures on the outskirts of the cluster \citep{2007A&A...470..425B, 2015Natur.528..105E, 2024A&A...692A.200G}. If star-forming galaxies in clusters originate from infalling field galaxies, their spatial distribution may align with the filaments or the projected cosmic web structures surrounding the cluster. Therefore, identifying star-forming galaxies in clusters may open a window to studying the cluster environment.

However, selecting star-forming galaxies in clusters is challenging. The spectral energy distributions (SEDs) of red galaxies usually exhibit characteristic features, such as the Balmer break, D4000, and 1.6 $\mu$m peak, while the SEDs of blue galaxies are significantly flatter. Strong emission lines, the bright intracluster light (ICL), and the blending of nearby galaxies can also lead to misleading photometric redshift results. Ideal surveys involve spectroscopic observations that cover the entire cluster \citep[e.g., the upcoming spectroscopic redshift survey project CHANCES, The Chilean Cluster Galaxy Evolution Survey,][]{2024arXiv241113655S}. However, this approach is very time-consuming, 
and requires large multi-object spectroscopic capabilities and a large field of view (FoV e.g., slitless spectroscopic survey by \citealt{2016ApJ...833..178V}; VLT/MUSE datacube by \citealt{2020A&A...644A..39D}; Anglo-Australian Telescope/AAOmega MOS by \citealt{2011ApJ...728...27O}.)

PAHs are large molecules (or small dust grains) and commonly serve as a dust extinction-free SFR indicator for massive star-forming galaxies \citep[see ][for a review]{2020NatAs...4..339L}. The PAH 3.3$\mu$m emission are sensitive to the massive star forming galaxies \citep[][]{2012ApJ...760..120K, 2020ApJ...905...55L, 2025A&A...693A.204V}. As a dust free SFR indicator, 3.3$\mu$m PAH emission traces star formation on timescales of approximately 3-10 Myr \citep{2024arXiv241207862J, 2025A&A...693A.204V}, making it a valuable probe of recent star formation activity. On the other hand, other observations also show that PAH features correlate well with the millimeter CO luminosity \citep{2024A&A...685A..78S, 2024arXiv241005397C, 2019MNRAS.482.1618C}, or AGN activity \citep{2012AJ....143...49W, 2019PASJ...71...25K, 2025arXiv250216301G}. Detailed spectroscopic studies of the Aromatic infrared bands (AIBs) shown that the fluxes and ratio between PAH 3.3 $\mu$m and the 3.4 $\mu$m aliphatic feature are sensitive to the intensity of the UV radiation field or neutral gas content \citep{2024A&A...685A..78S, 2025arXiv250218464}. 

In this work, we make use of the medium band filter F430M to select the 3.3 $\mu$m PAH emitter to trace the star formation in galaxy cluster A2744. The JWST images, which recently covered the entire cluster through various projects, are deep enough to allow us to select emission-line galaxies in and around the cluster. The galaxy cluster A2744 at $z = 0.3$ has the PAH 3.3 $\mu$m emission line shifted to the F430M band. The 5$\sigma$ depth of the F430M and F444W images is about 27.72 and 29.2 AB mag in primary field, and 27.07 and 28.8 AB mag in the Parallel field, respectively \citep{2024ApJ...976..101S, 2024ApJ...974...92B}, 
providing an excellent sample of 3.3 $\mu$m PAH emitters down to a SFR$_{\rm SED} \gtrsim 1 M_\odot \rm yr^{-1}$, facilitating the exploration of the latest star-forming stages of the galaxy cluster before quenching.

To study the star formation activity in galaxy clusters, we select a sample of 3.3$\mu$m PAH bright galaxies from the F430M JWST/NIRCam image. The F430M image covers the A2744 as well as the JWST/NIRISS Parallel region, thus the star-forming galaxies selected from the F430M imaging allow us to explore a wide cluster's area, including galaxies as far as \citep[$ \sim 1\times R_{200} =  2.0$ Mpc or $7.3'$, ][]{2014A&A...562A..11I} from the centre, distances that start mixing to field galaxies from the control sample \citep{2015ApJ...806..101H, 2024MNRAS.527L..19L}. Meanwhile, the high resolution of the JWST imaging enables us to study PAH morphology, which can be used to measure the surface density of SFR. Additionally, the lopsidedness of the PAH morphology can help us connect the ram pressure strength with gas and star formation properties.

On the other hand, A2744 has a wealth of archival datasets, including imaging data from HST and JWST ranging from 0.4 to 5 $\mu$m, spectral data from VLT/MUSE and JWST/NIRSpec \citep{2020A&A...644A..39D, 2024arXiv240803920P}. Specially, A2744 is covered in the Herschel Lensing Survey \citep[HLS,][]{2010A&A...518L..12E}, and will be helpful to calibrate the SFR$_{\rm PAH}$ in clusters, as well as connect the PAH and hot dust continuum properties with the cold dust mass \citep{2018A&A...620A.125M}. Given the extensive data available, the dark matter halo is well-modeled by several studies \citep{2011MNRAS.417..333M, 2023ApJ...952...84B, 2023MNRAS.523.4568F, 2024arXiv241019907C}, which aids in linking the mass density of the environment with star formation. 

Throughout the paper, we adopt the Chabrier IMF \citep{Chabrier2003} and the standard Lambda cold dark matter cosmology ($\Lambda$CDM) with $\rm \Omega_m=0.3$, $\rm \Omega_\Lambda=0.7$, and H$\rm _0=70 km\,s^{-1}\,Mpc^{-1}$, 
and the AB magnitude system \citep{okeSecondaryStandardStars1983}.

\begin{figure}[ht]
    \centering
    \includegraphics[width=0.95\linewidth]{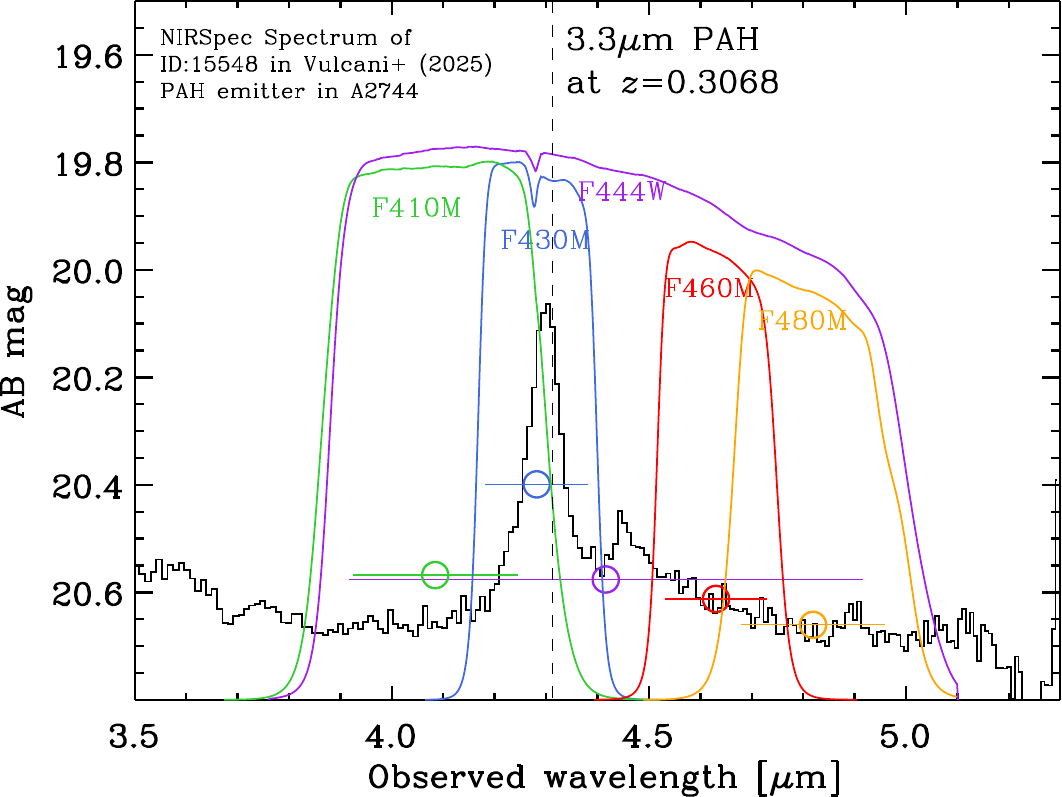}
    \caption{Transmission curves of JWST/NIRCam filters F410M (green), F430M (blue), F460M (red), F480M (orange), and F444W (purple), overlaid with the NIRSpec spectrum of one PAH emitter in A2744 \citep[ID: 15548 in ][]{2025A&A...693A.204V} at redshift $z = 0.3068$ (PI M. Castellano, GO-3073). Open circles indicate the observed flux of the spectrum through each filter. The F430M filter (blue) is uniquely positioned to encapsulate the redshifted 3.3 $\mu$m PAH emission (dashed line at 4.3 $\mu$m), while the adjacent filters (F410M, F460M, F480M) sample the stellar continuum. The F444W filter has the most overlap in imaging coverage with F430M, making F430M and F444W the chosen bands for selecting sources based on PAH emission and continuum. This target is the ID 5565 in this work.
    }
    \label{filter}
\end{figure}

\begin{figure}[ht]
    \centering
    \includegraphics[width=0.99\linewidth]{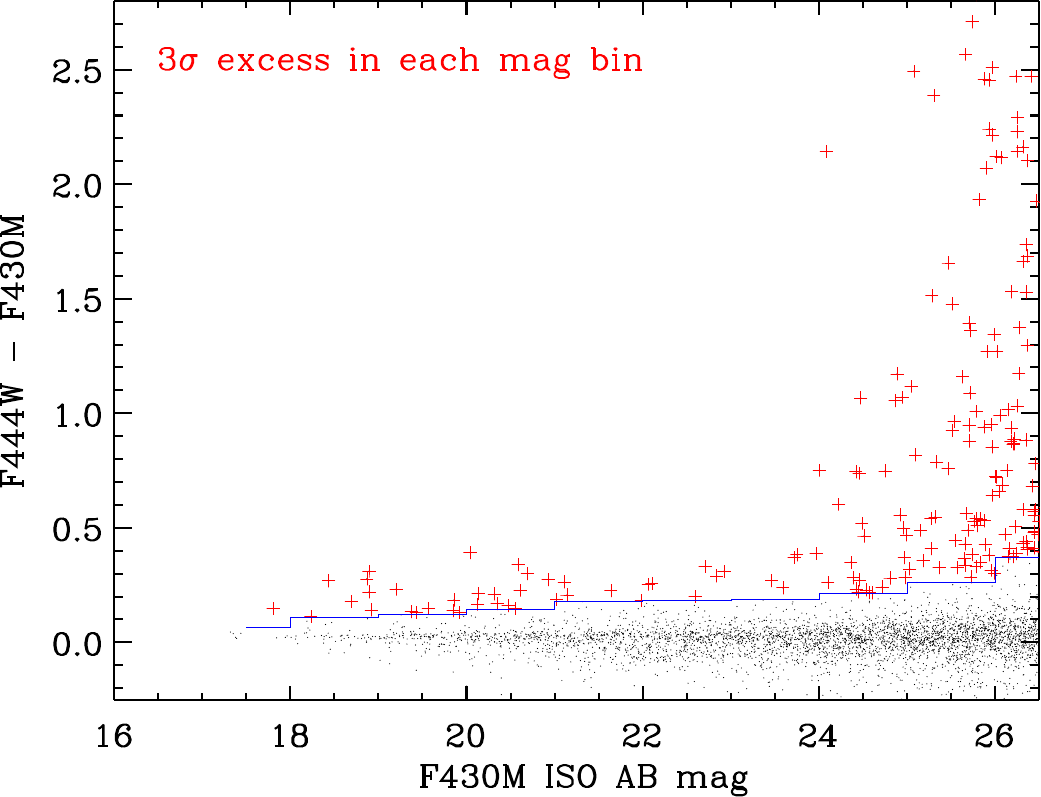}
    \includegraphics[width=0.99\linewidth]{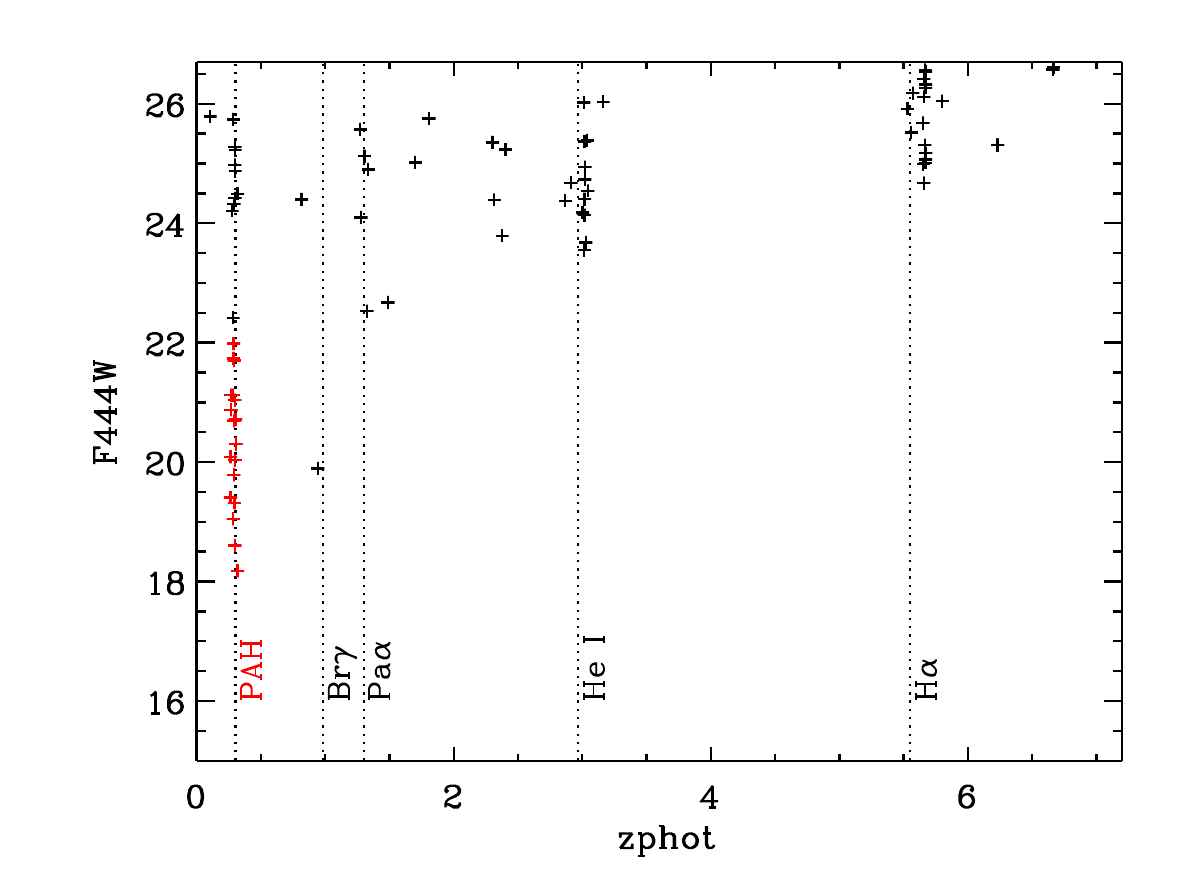}
    \caption{
    {\bf Upper panel:} F430M-F444W color versus F430M iso mag of the F430M selected sample. The blue line marks the $3\sigma$ detection limit of the color excess. The $3\sigma$ excess F430M emitters are highlighted with red plus signs. 
    {\bf Lower panel:} Photometric redshift distribution of the F430M emitters. The photometric redshifts are separated into several redshift bins, corresponding to several emission lines that shifts to F430M filter. The brightest targets are mainly the 3.3$\mu$m PAH emitters. We highlight the PAH targets studied in this work by red crosses. 
    }
    \label{f430m}
\end{figure}

\begin{figure*}[ht]
    \centering
    \includegraphics[width=0.95\linewidth]{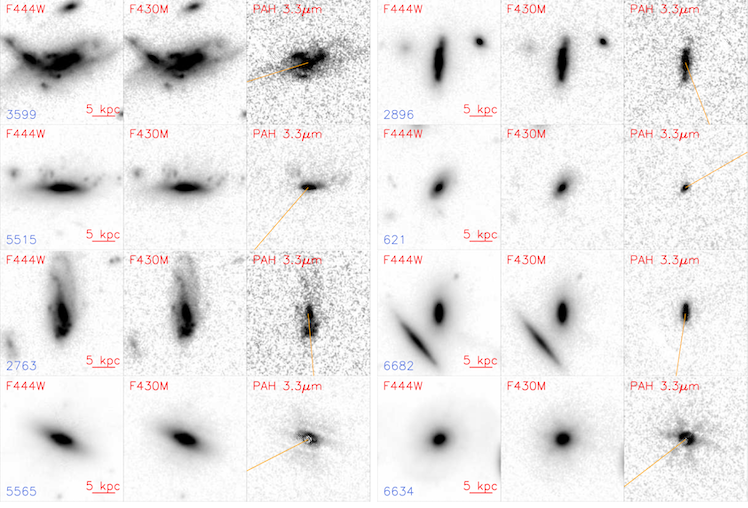}
    \caption{
    Stamp images of the 3.3$\mu$m PAH bright targets ($6''\times 6''$). We show the F444W, F430M and the PAH 3.3$\mu$m images (=F430M - F444W) of each target. The IDs are denoted in the left corner. The orange line in the PAH images show the projected direction to the cluster center \citep[RA = 00:14:20.7022; Dec = -30:24:00.6264,][]{wangGRISMLENSAMPLIFIEDSURVEY2015,2018MNRAS.473..663M,2023ApJ...952...84B}. The clear detection in PAH images confirms our selection method. PAH images reveal complex morphologies, including disky, clumpy, or compact structures. Additionally, some exhibit asymmetric features, suggesting the influence of ram pressure.
    }
    \label{stamp1}
\end{figure*}

\begin{figure*}[ht]
    \centering
    \includegraphics[width=0.9\linewidth]{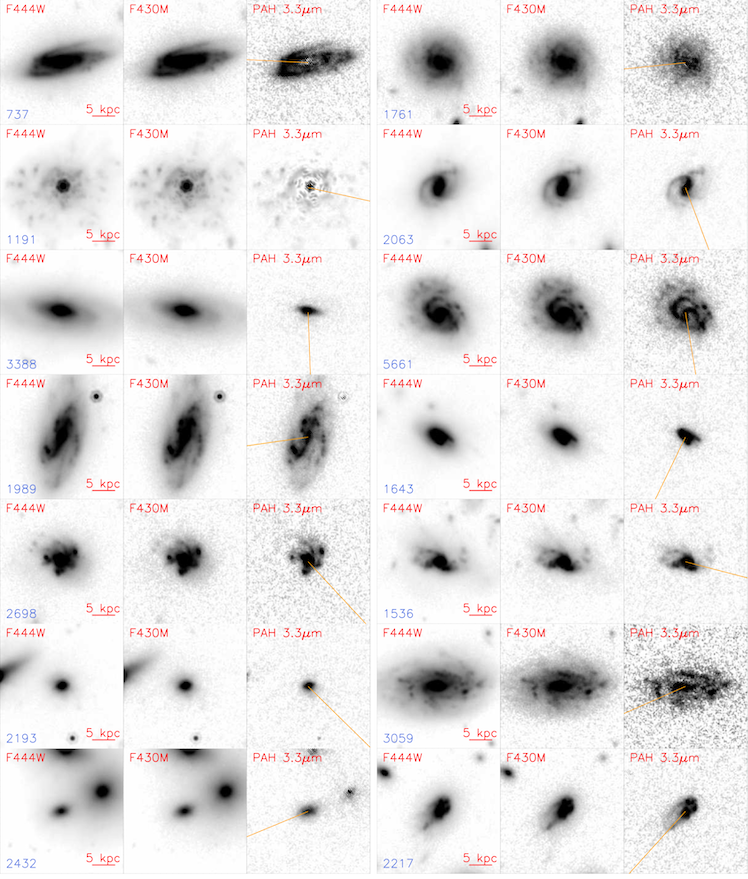}
    \caption{
    Same caption as Figure \ref{stamp1}.
    }
    \label{stamp2}
\end{figure*}

\begin{figure*}[ht]
    \centering
    \includegraphics[width=0.99\linewidth]{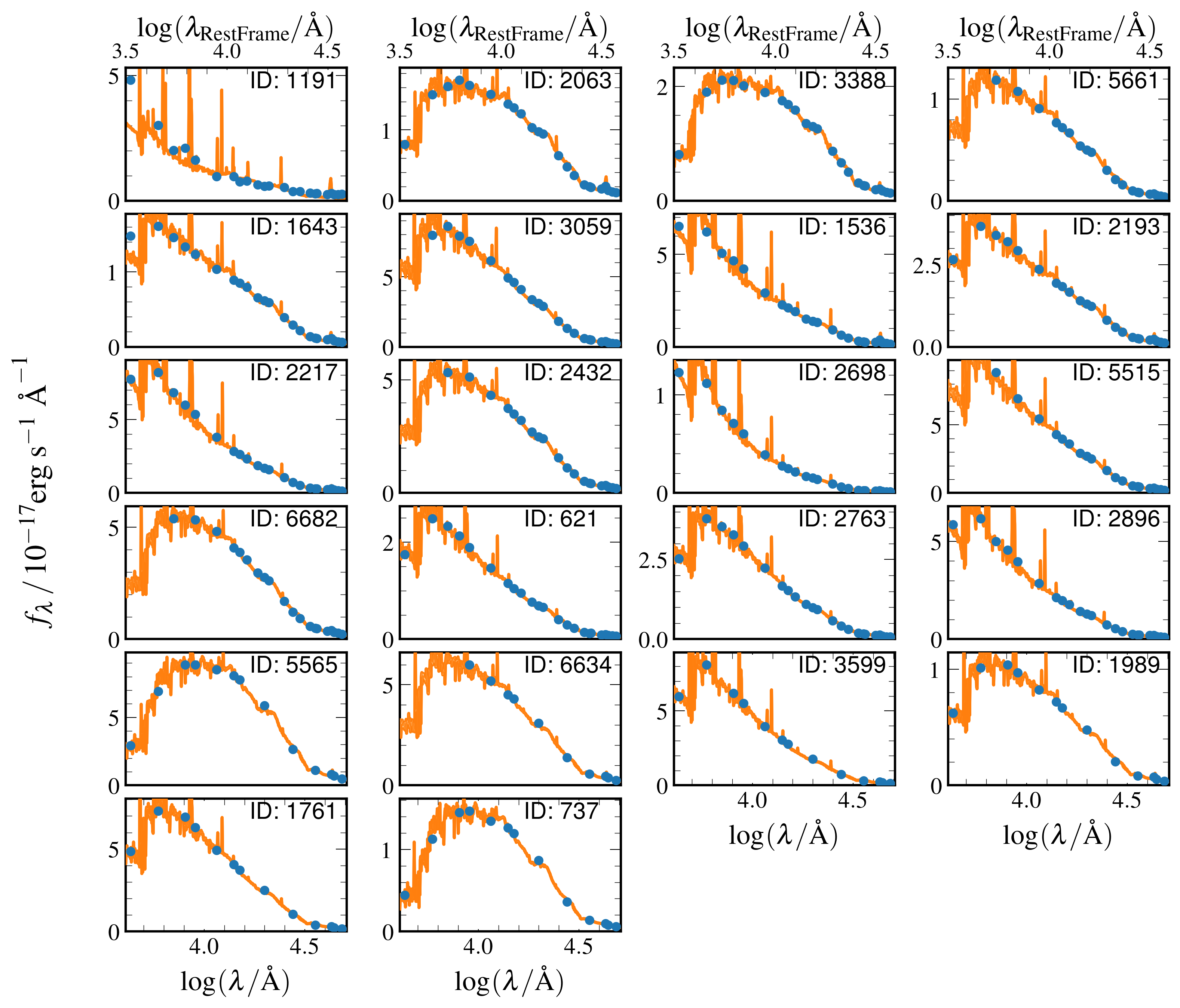}
    \caption{The Bagpipes SED fitting results for the PAH emitter sample. The orange curves represent the model spectra, and the blue dots show the photometric data points from the SED released by UNCOVER \citep{2024ApJS..270....7W}. The blue end of ID 1191 is not well fitted due to the central AGN contribution \citep{2012ApJ...750L..23O}. A more detailed SED analysis of this target is presented by \citet{2024arXiv240915215W}.
    }
    \label{SED}
\end{figure*}

\begin{figure*}[ht]
    \centering
    \includegraphics[width=0.99\linewidth]{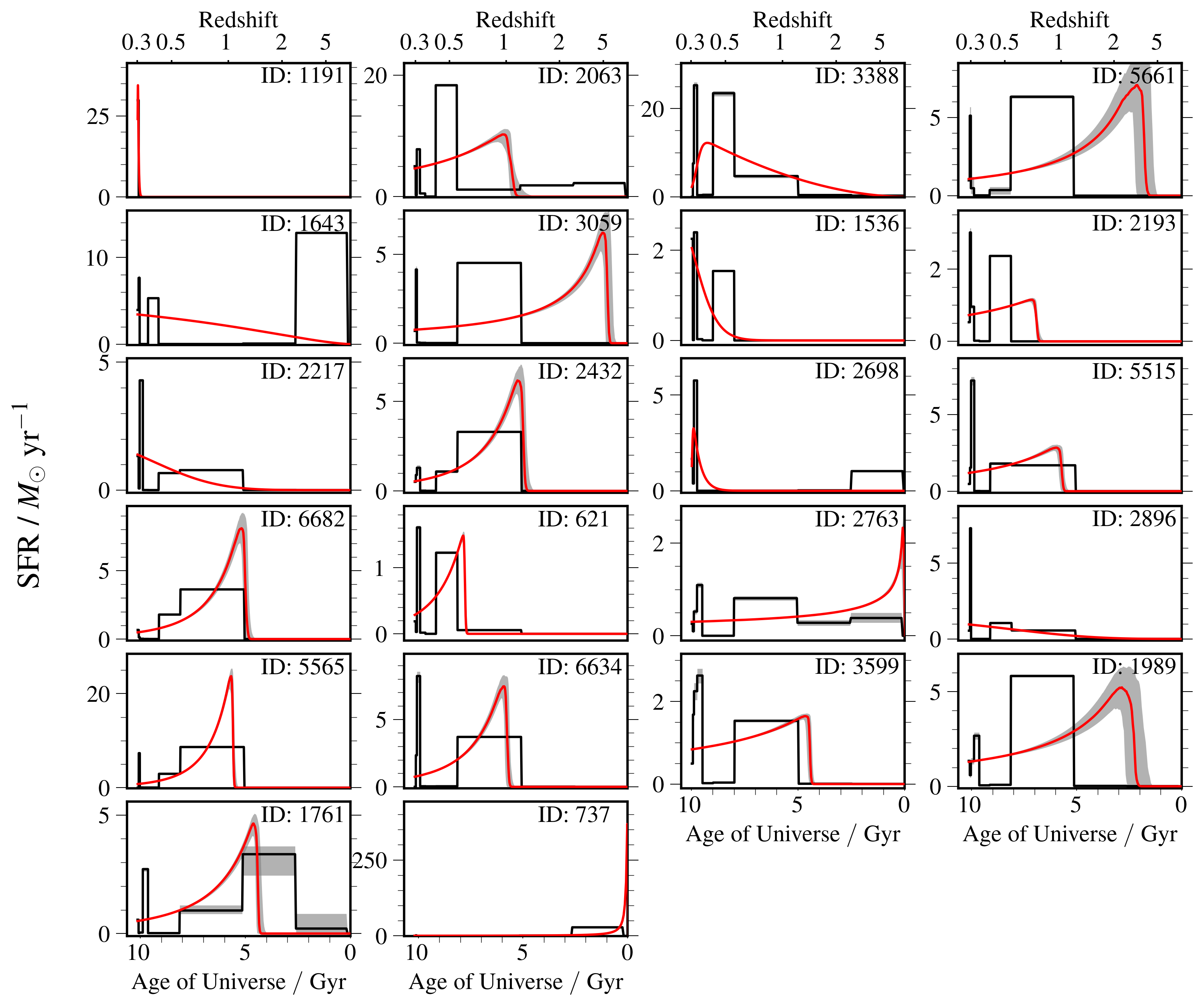}
    \caption{Star formation history derived from Bagpipes SED fitting. The red curves represent the double power-law SFH, while the black stepped curves correspond to the continuity non-parametric SFH model results. The two SFH models are generally consistent, though the non-parametric SFH reveals a recent starburst feature that is smoothed out in the double power-law model. Most of the stellar mass formed around redshift 1.
    }
    \label{SFH}
\end{figure*}

\begin{figure}[ht]
    \centering
    \includegraphics[width=0.95\linewidth]{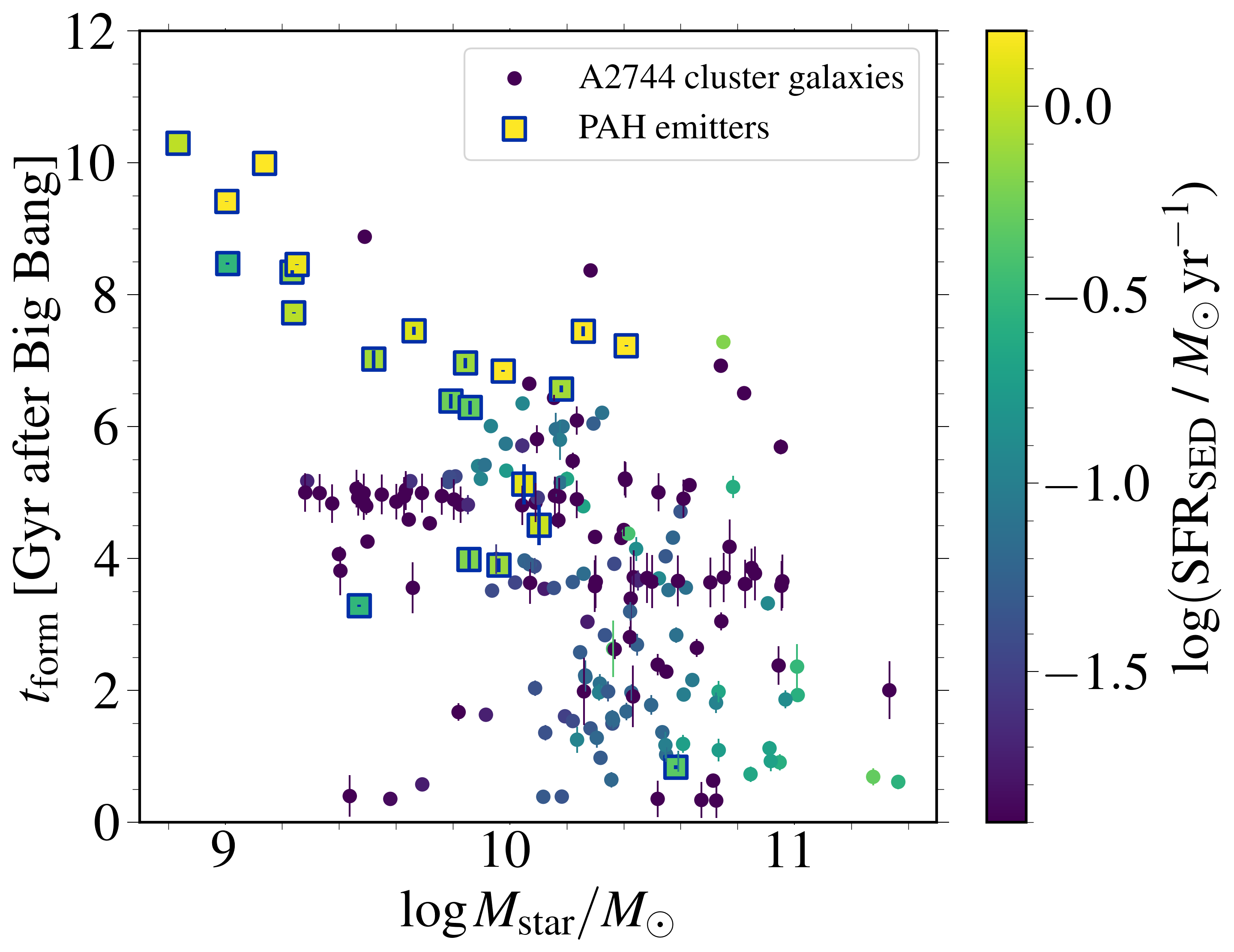}
    \caption{
    Stellar mass and the mass-weighted formation timescale $t_{\rm form}$ distribution. The cluster galaxies (dots) have a mass-weighted formation time of 4 Gyr after the big bang.    The PAH sample (squares) are formed more recently. The color bar shows the star formation rates, which are higher for the PAH sample and the most massive cluster galaxies. 
    }
    \label{tform}
\end{figure}

\begin{figure}[ht]
    \centering
    \includegraphics[width=0.95\linewidth]{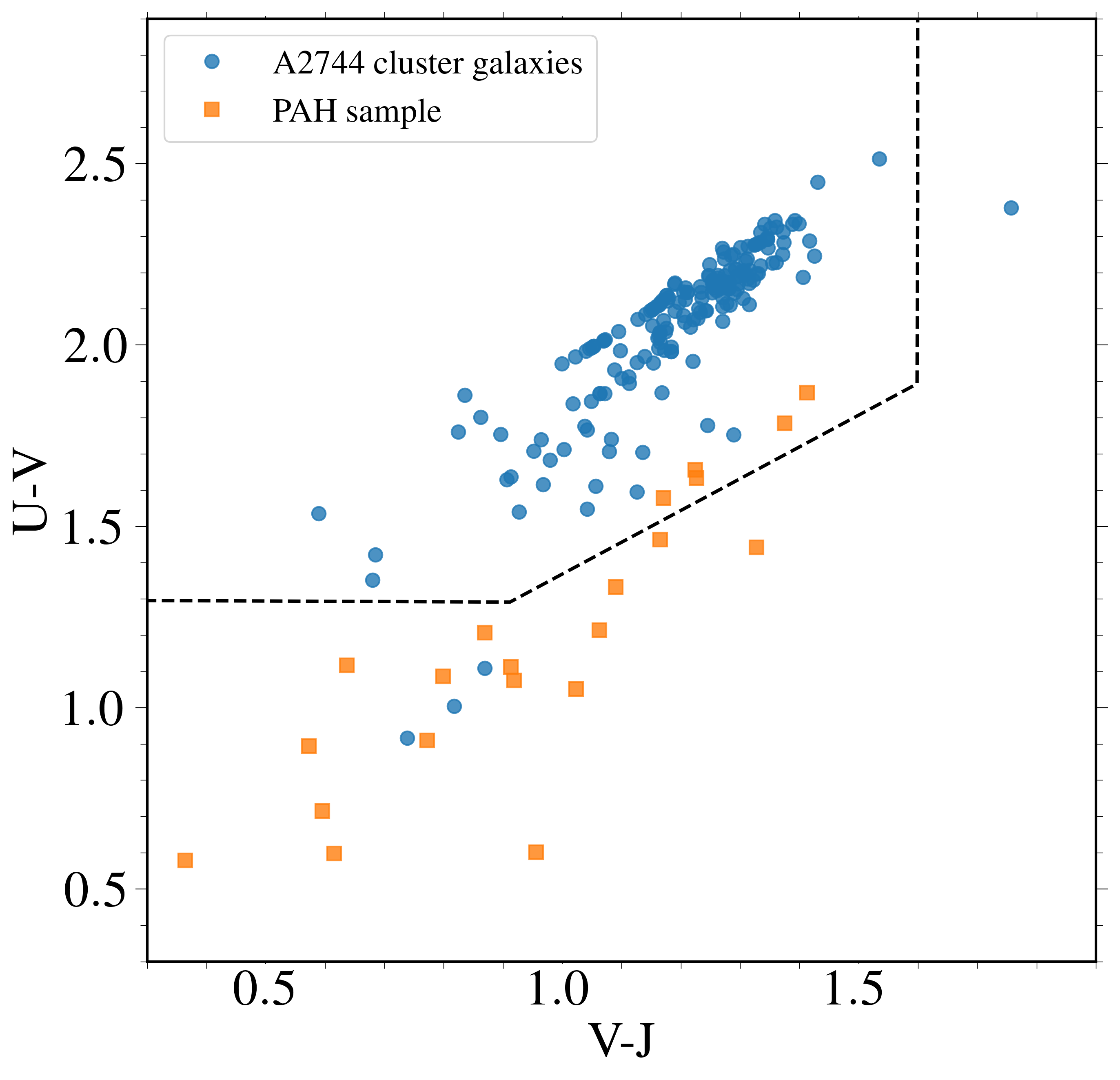}
    \caption{The U-V v.s. V-J distribution of the cluster galaxies (blue dots) and PAH sample (orange square). The dashed line are the critical lines to divide the quiescent and star forming galaxies \citep{2018ApJ...858..100F}. So most of the cluster members is quiescent, and the PAH sample are mainly star forming.
    }
    \label{sfr}
\end{figure}

\begin{figure*}[ht]
    \centering
    \includegraphics[width=0.8\linewidth]{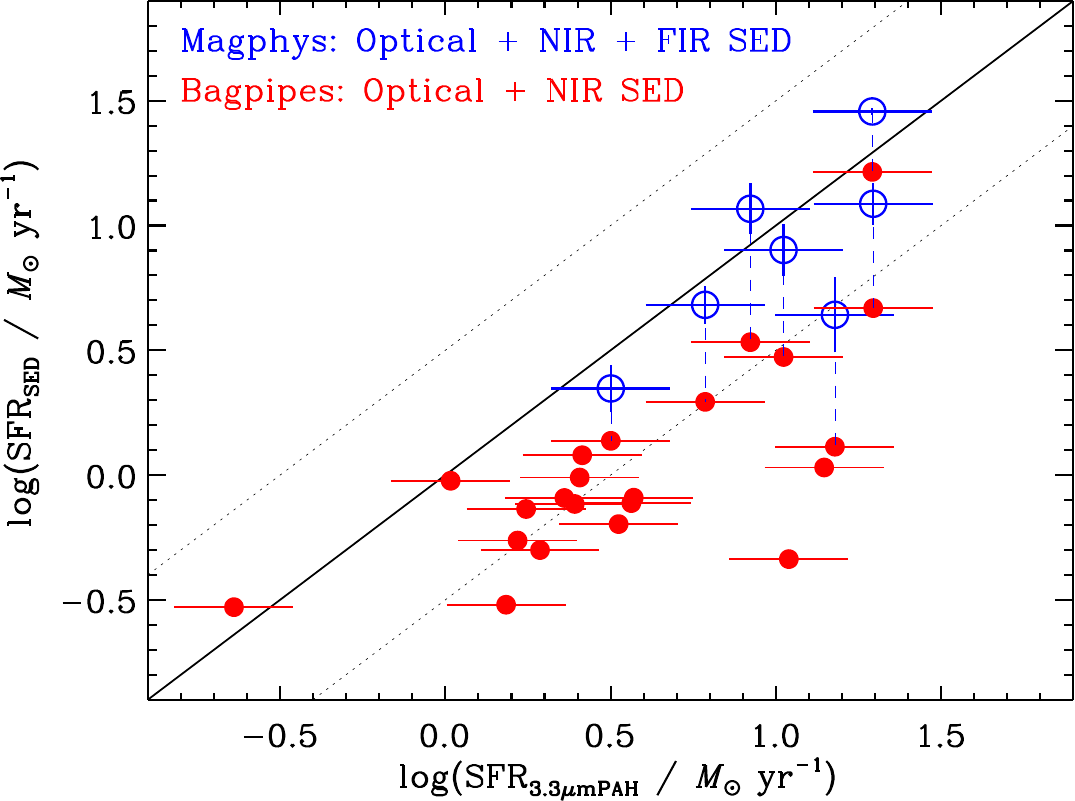}
    \caption{
    Comparison of SFR$_{\rm 3.3\mu mPAH}$ with SFR$_{\rm SED}$ estimates from SED fitting using Bagpipes and Magphys. Red points indicate SFR$_{\rm SED}^{\rm Bagpipes}$, derived from Optical-to-NIR SED fitting, which primarily traces star formation from the stellar population. Blue open circles represent SFR$_{\rm SED}^{\rm Magphys}$ of the seven Herschel detected PAH emitters.  SFR$_{\rm SED}^{\rm Magphys}$ values are obtained from Optical-to-FIR SED fitting by Magphys, incorporating dust emission and thus providing a more complete measure of total star formation. The solid black line denotes the 1:1 relation, with dashed black lines indicating a $\pm$0.5 dex range. For the seven Herschel-bright targets, SFR$_{\rm SED}^{\rm Magphys}$ systematically exceeds SFR$_{\rm SED}^{\rm Bagpipes}$ and is in better agreement with SFR$_{\rm 3.3\mu mPAH}$ (blue circles). The blue dashed lines connect the two SFR$_{\rm SED}$ estimates for the same source, illustrating the difference between the two methods. The overall consistency suggests that the 3.3$\mu$m PAH flux, estimated from medium-band photometry, remains a reliable SFR tracer, immune to dust obscuration.
    }
    \label{SFRPAH}
\end{figure*}

\begin{figure*}[ht]
    \centering
    \includegraphics[width=0.8\linewidth]{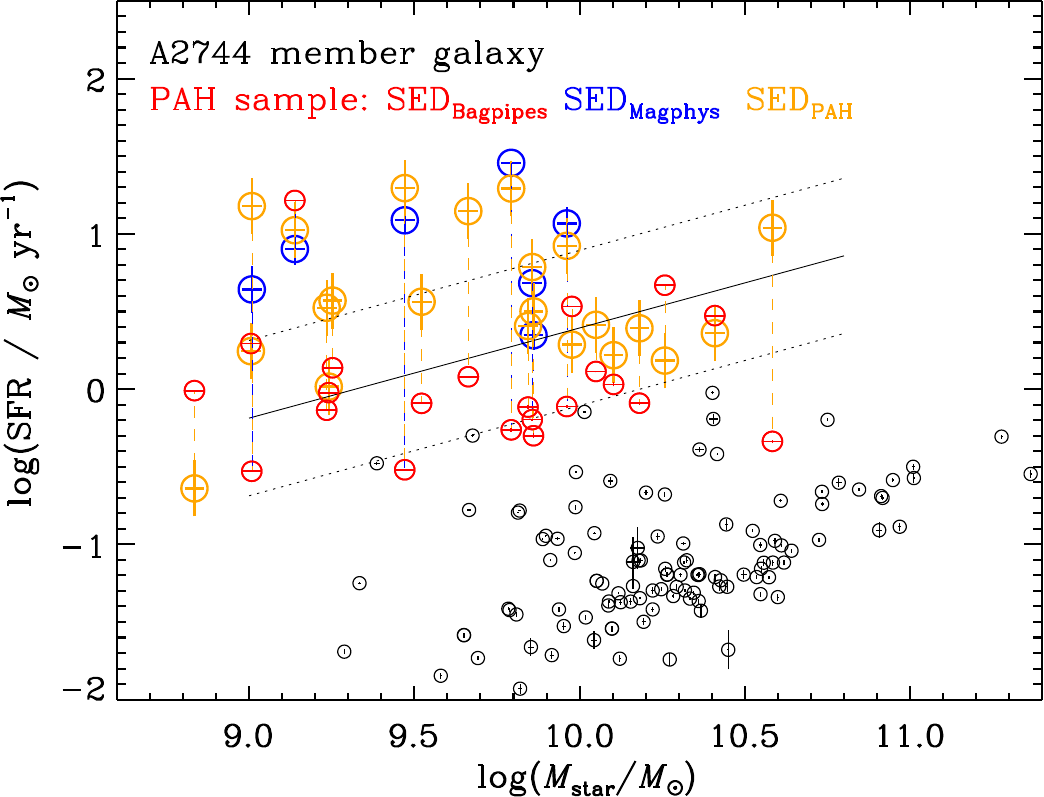}
    \caption{The star-forming main sequence of the PAH targets (orange circles for SFR$_{3.3\mu m \rm PAH}$, blue circles for SFR$_{\rm Magphys}$, and red circles for SFR$_{\rm Bagpipes}$) and the cluster member galaxies (black open dots). We link the SFR for the same target with dashed lines. 
    Most of the cluster galaxies are quiescent, and therefore have low SFR. We only show cluster galaxies with SFR above $0.01 M_\odot\,\rm yr^{-1}$, which is enough to show the separation between the PAH sample and the SFR of other cluster members.
    The solid and dotted lines shows the star forming main sequence \citep{2014ApJS..214...15S} at $z = 0.308$ and a scatter of 0.5 dex. The PAH bright targets have a similar or higher star formation rate as the field star forming galaxies. Most of the cluster galaxies have the star formation rate one order of magnitude lower than the star forming galaxies. So the 3.3 $\mu$m PAH selection targets are mainly on or above the main sequence. 
    }
    \label{ms}
\end{figure*}

\section{Sample selection and archival data}

\subsection{Archival Data in A2744}

A2744 is one of the six clusters observed by the Hubble Space Telescope Frontier Fields project \citep{2017ApJ...837...97L}, reaching a point-source 5$\sigma$ depth of $\sim 29$ AB mag in F435W, F606W, F814W and F105W, F125W, F140W, F160W. VLT/MUSE observation of this field provide spectroscopic redshift results deep to about $m_{\rm 1500}\sim 30$ mag \citep{2020A&A...644A..39D}. This field is also covered by Herschel and ALMA in far infrared and submillimeter bands \citep{2016MNRAS.459.1626R, 2017A&A...597A..41G, 2022ApJ...932...77S, 2022ApJS..263...38K, 2023arXiv230907834F}, as well as X-ray observations \citep{2014A&A...562A..11I, 2015Natur.528..105E, 2024A&A...692A.200G}. After the launch of JWST, A2744 has been observed by JWST/NIRCam and NIRISS in F070W, F090W, F115W, F140M, F150W, F158M, F162M, F182M, F200W, F210M, F250M, F277W, F300M, F335M, F356W, F360M, F410M, F430M, F444W, F460M, F480M bands, as well as JWST/NIRSpec spectroscopic observations \citep[][also DDT-2856, GO-2883, GO-3538]{wangEarlyResultsGLASSJWST2022,2024ApJ...976..101S, heEarlyResultsGLASSJWST2024, jiangLyaNondetectionJWST2024, 2024ApJ...974...92B, 2024arXiv241001874N, liEarlyResultsGLASSJWST2025}. All the HST and JWST images can be downloaded from the UNCOVER \citep{2024ApJ...974...92B, 2024ApJ...976..101S} website\footnote{\url{https://jwst-uncover.github.io/DR3.html\#Mosaics}}. The deep and high resolution images from HST and JWST, and the deep spectroscopic redshift surveys made A2744 one of the best deep fields for extragalactic cluster studies.

\subsection{PAH Sample Selection}

We show the NIRCam filter F410M, F430M, F460M, F480M and F444W filter response curve \citep{2023PASP..135b8001R} in Figure \ref{filter} with the 
spectrum of one PAH bright target in A2744 \citep[ID: 15548 in ][]{2025A&A...693A.204V} taken by JWST/NIRSpec (PI M. Castellano, GO-3073) as a template of PAH bright galaxy. The spectrum is obtained from DAWN JWST Archive \citep[DJA,][]{2023zndo...8319596B, 2024arXiv240905948D, 2024Sci...384..890H}. 
At the redshift of $z=0.3$, the 3.3 $\mu$m PAH shifts to F430M filter, and the F410M, F460M, F480M flux probing the continuum emission. The F444W flux includes the PAH emission and the continuum, and approximately close to the continuum flux because of the broad wavelength coverage (FWHM of F444W is 11144.5\AA). And since the F444W filter has the widest overlap with the F430M coverage, we use the F430M and F444W images as the emitter band and continuum band, respectively. Since the F444W band image includes the 3.3 $\mu$m PAH emission, using F430M + F444W imaging to detect PAH emitters represents a compromise between survey area and an accurate continuum estimate. We will discuss the potential bias in Section \ref{bias}.

We perform photometry using SExtractor \citep{1996A&AS..117..393B} in dual mode, with the F430M image for detection and the F444W image for photometry. We use the ISO magnitude for the target selection, which is the aperture with all the high F430M signal-to-noise pixels to optimize target selection based on the F430M excess. Then we obtain a total of 76 candidates with a $3\sigma$ excess in the F430M band (Figure \ref{f430m}). After identifying the F430M emitters, we cross-match our sample to the psf-matched multi-wavelength catalog provided by UNCOVER \citep{2024ApJS..270....7W}, taking into account the results of photometric redshifts (phot-z) and spectral energy distribution (SED) modeling covering the wavelength range from F435W to F444W bands \citep{2024ApJS..270...12W}. 

We show the phot-z and F444W magnitude in Figure \ref{f430m}. As expected, these F430M emitters are distributed in several redshift bins, including $z = 0.3$ for PAH, $z = 1.6$ for Pa$\alpha$, $z = 3$ for He I ($\lambda = 10833\rm \AA$), $z = 5.5$ for H$\alpha$ \citep{2024arXiv240810980M}. The low redshift of the cluster and the wide wavelength coverage in rest frame enable a reliable photo-z estimation to exclude emission line targets at other redshifts, resulting in PAH-selected galaxies at the redshift of A2744 (Figure \ref{f430m}, at $z = 0.3$). 

Observations have shown that the PAH emission in dwarf galaxies is faint \citep{2004ApJS..154..211H, 2005ApJ...624..162H}, so for this reason we decided to focus on the properties of the massive star-forming galaxies in A2744 by applying a cut in the F444W magnitude at F444W $<$ 22 AB mag, corresponding to roughly $M_*/M_\odot \simeq 10^{9}$ (CANDELS catalog, Figure \ref{mass}). We will discuss the potential biases of the flux cut in Section \ref{bias}. The final sample includes 22 PAH bright targets selected from the F430M and F444W images (Table \ref{tab}). We show the stamp images in Figure \ref{stamp1} and Figure \ref{stamp2}. 

To compare the star formation properties of this PAH bright sample, we also select targets with F200W $ < 19$ and $0.28 < z_{\rm spec} < 0.34$ \citep[based on the redshift distribution presented in][]{2011ApJ...728...27O} from UNCOVER catalog, which represents the brightest targets in A2744 cluster.

\section{Analyze Results} \label{sec:results}

\subsection{SED fitting}

The SED fitting analysis is performed using Bagpipes \citep{2018MNRAS.480.4379C, 2019MNRAS.490..417C}, on photometric catalog from UNCOVER \citep{2024ApJS..270....7W}, covering the wavelength range 
from 0.435 to 4.8 $\mu$m.
The filters used include HST bands (F435W, F606W, F814W) and JWST bands (F070W, F090W, F115W, F140M, F150W, F162M, F182M, F200W, F210M, F250M, F277W, F300M, F335M, F356W, F360M, F410M, F430M, F444W, F460M, F480M).
Leveraging advanced Bayesian inference techniques and flexible model configurations, Bagpipes enables precise and rapid fitting of complex processes. We use the default parameters including the double powerlaw SFH, Chabrier IMF \citep{Chabrier2003}, ionization parameter $\log U = -3$, and the attenuation curve by \citet{calzettiDustContentOpacity2000}. Nine targets have spectroscopic redshifts, which are adopted in Bagpipes, and the rest targets were set at $z = 0.308$.

The Bagpipes fitting results are shown in Figure \ref{SED}. The high S/N and the wide wavelength coverage ensure a reliable constrain of the stellar properties. The blue end in the SED of the target ID 1191 is not fitting well because of the central AGN (Section \ref{target}). To assess the robustness of the SED fitting, we also applied the continuity non-parametric star formation history (SFH) model \citep{2019ApJ...876....3L} using Bagpipes. To better characterize recent star formation activity, we adopted time bins of [0, 20, 50, 100, 250, 500, 1000, 2000, 5000, 7500, 10000] Myr in the continuity model, with the results shown in Figure \ref{SFH}. The overall trends in the two SFH reconstructions are consistent. However, the non-parametric fitting generally reveals a recent starburst peak around $z\simeq 0.3$, in agreement with the star formation activity indicated by PAH emission. Most of the stellar mass formed at $z \simeq 1$.

We present the mass-weighted formation timescale quantitatively in Figure \ref{tform}. As expected, the F444W flux limit leads to the stellar mass higher than $10^{9} M_\odot$. While for the other galaxies in A2744, the stellar mass can be as high as $10^{11.5} M_\odot$, which are mainly the massive quiescent galaxies in clusters. The formation time $t_{\rm form}$ of the galaxies are mainly 4 Gyrs after Big Bang or earlier, which is consistent with the age of the ICL in this merging cluster \citep{2014ApJ...794..137M}. On the other hand, the PAH selected sample are formed more recently. Consequently, the mass-weighted age of the PAH bright sample is the youngest among the cluster member populations, and thus the PAH sample traces the most recent star formation activity within the galaxy cluster. We also show the UVJ diagram in Figure \ref{sfr}. As expected, the PAH emitters are mainly star forming. 

SFRs from SED fitting are highly dependent on the assumption of SFH. In Figure \ref{SFR_SFH}, we verify the SFR$_{\rm SED}$ results from different SFH.

Our 3.3$\mu$m selected sample can also include the dusty star-forming galaxies, which would also be detected in far infrared bands. We cross match our sample with the Herschel Lensing Survey catalog \citep[][]{2010A&A...518L..12E, 2016MNRAS.459.1626R} within 10 arcsec, and matched 7 targets with Herschel detection. The maximal distance between the PAH sample and corresponding Herschel targets are lower than 0.3 arcsec. Thus the optical counterparts of the far infrared targets are reliable, despite the 50 times different resolution. Moreover, two PAH emitters (ID 2063 and 1643) are also detected by ALMA in the DUALZ project \citep{2023arXiv230907834F}, with DUALZ catalog IDs of 68 and 16, respectively.

For the seven Herschel detected targets, we add the Herschel SED in PACS 100$\mu$m, 160$\mu$m bands, and SPIRE 250$\mu$m, 350$\mu$m, 500$\mu$m, and fit the SED with MAGPHYS \citep{2008MNRAS.388.1595D}. MAGPHYS can estimate the SFR$_{\rm SED}$ based on UV and FIR emission based on the assumption that the dust extincted UV photon energy would be re-radiated into FIR bands, and thus provide the $total$ SFR results. 

\begin{figure*}[ht]
    \centering
    \includegraphics[width=0.33\linewidth]{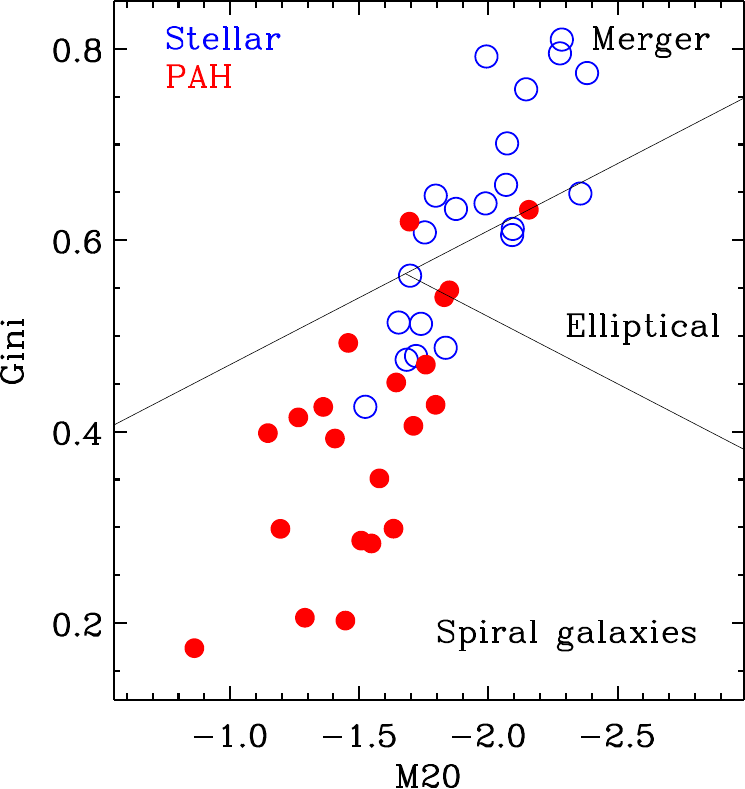}
    \includegraphics[width=0.327\linewidth]{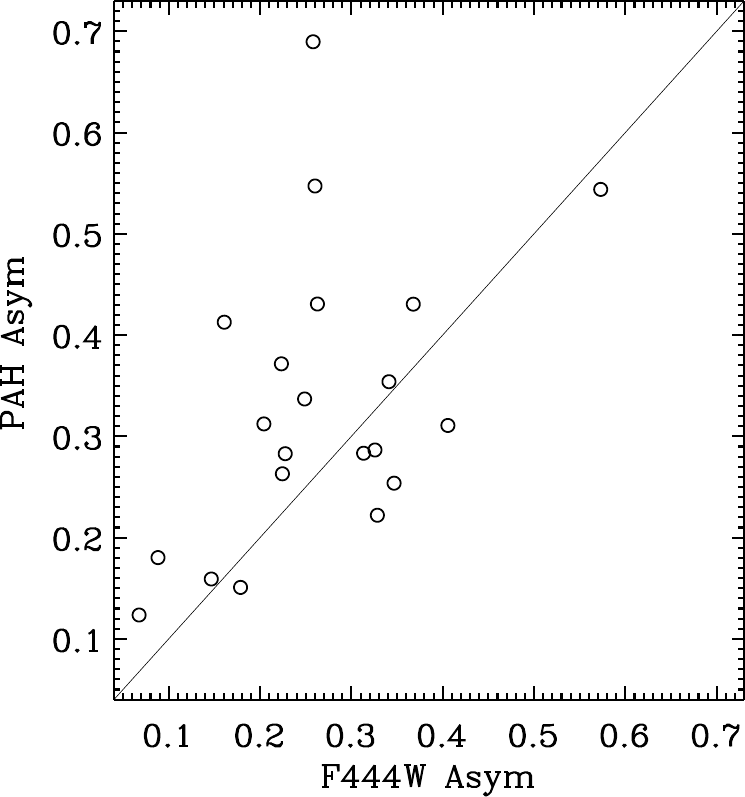}
    \includegraphics[width=0.33\linewidth]{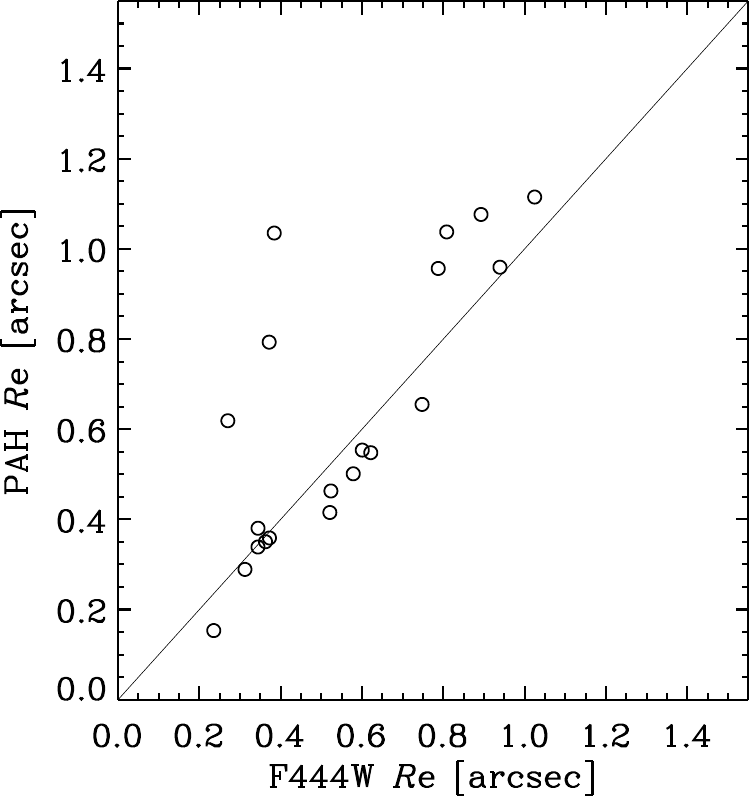}
    \caption{
    Morphology parameters of the PAH distribution (from the F430M - F444W image) and the stellar distribution (from the F444W image).
    {\bf Left:} Gini-M20 results with the reference lines from \citet{2004AJ....128..163L}. The PAH morphologies are close to the normal galaxy region, while the stellar distributions are close to ULIRGs region. The change from the stellar distribution to the PAH morphology is from compact (high Gini, low M20) to disky (low Gini, high M20).
    {\bf Middle:} Asymmetry of the targets of PAH and stars. The PAH distribution is more asymmetric than stellar distribution, especially for the targets with Asymmetry index $>0.4$.
    {\bf Right:} Half light radius of the stellar and PAH distribution. The PAH emission region is typically more extended than the stellar continuum. The PAH morphology parameters are measured from the F430M image with F444W image subtracted.
    }
    \label{morph}
\end{figure*}

\begin{figure}[ht]
    \centering
    \includegraphics[width=0.99\linewidth]{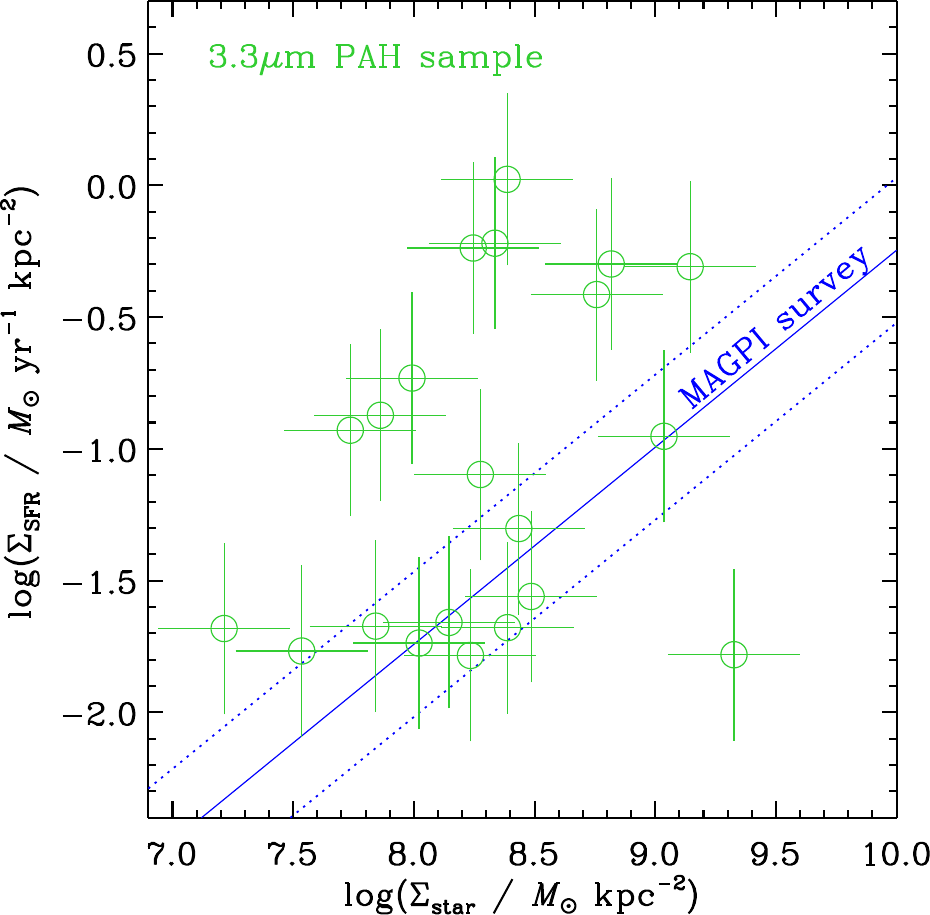}
    \caption{Star formation surface density vs. stellar surface density of the PAH bright sample (green open circles). The blue lines represent the scaling relation of star forming galaxies at $0.25<z<0.35$ from the MAGPI survey project \citep{2024MNRAS.530.5072M}. The star formation surface density of the PAH bright sample is higher than the scaling relation. The only exception is target ID 3599, located in the lower-right corner, which is blended with nearby galaxies (see Figure \ref{stamp1}). This blending affects the F444W size measurement, leading to an anomalously high stellar surface density.
    }\label{SigmaSFR}
\end{figure}

\begin{figure}
    \centering
    \includegraphics[width=0.99\linewidth]{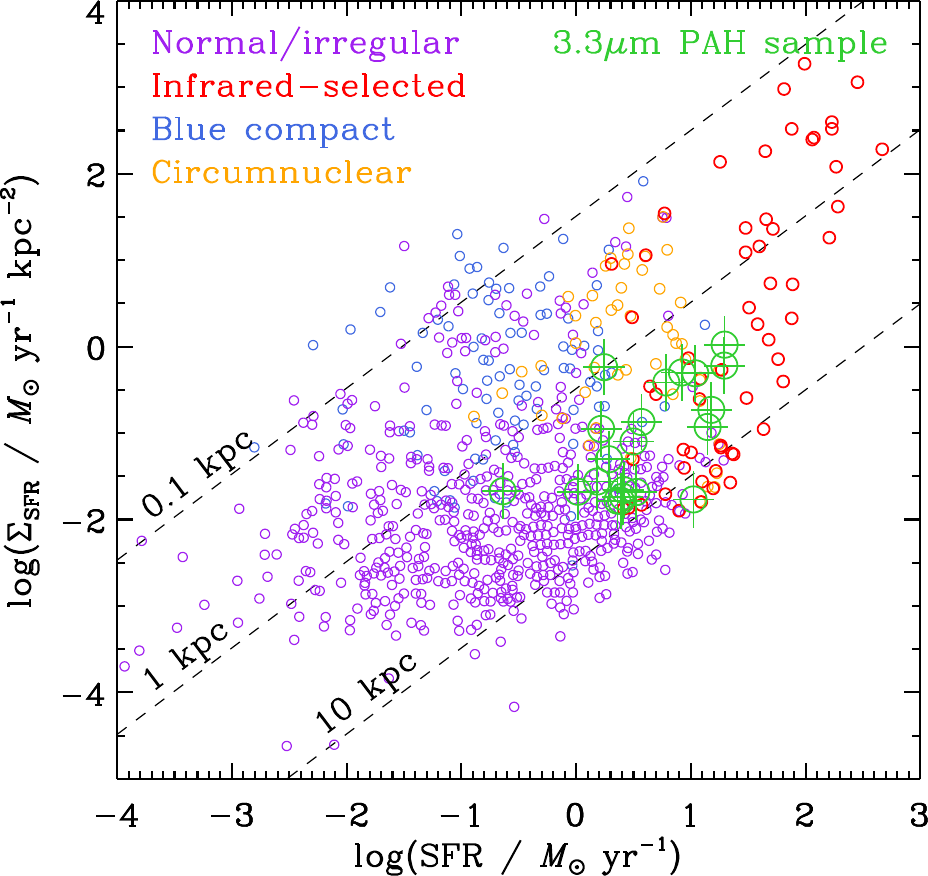}
    \caption{The star formation rate and star formation surface density distribution of the 3.3 $\mu$m PAH sample (green), as compared with normal/irregular galaxies (purple), infrared-selected galaxy (red), blue compact starburst galaxies (blue), and circumnuclear star-forming ring in local barred galaxies (orange) from \citet{2012ARA&A..50..531K}. The 3.3 $\mu$m PAH sample lies near the transition region between normal and infrared selected galaxies. The lack of compact star-forming galaxies with similar star formation rates may be due to the small sample size. Another potential bias of the PAH bright sample among the star forming population is the lack of the blue compact galaxies, which may due to the lack of PAH in dwarf galaxies (Section \ref{bias}). 
    }
    \label{SFRSigmaSFR}
\end{figure}

\begin{figure*}[ht]
    \centering
    \includegraphics[width=0.9\linewidth]{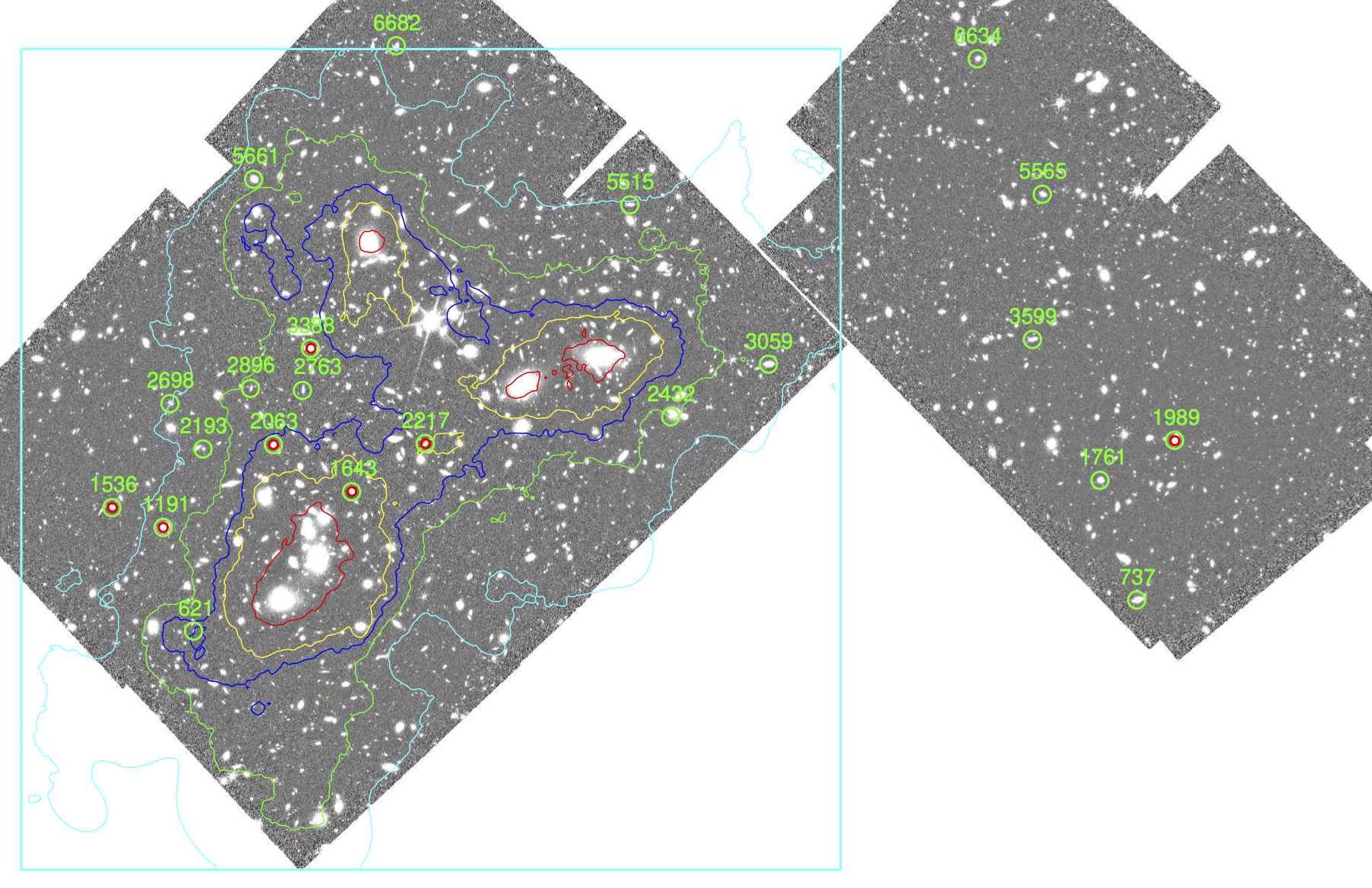}
    \caption{F430M image of A2744, with the 3.3$\mu$m PAH emitters highlighted. Targets within the red circles are the Herschel detected galaxies in A2744 cluster. The cyan contours are the mass surface density with levels of [2, 4, 6, 8, 16]$\times 10^8\, M_\odot\,\rm kpc^{-2}$ from \citet{2024ApJ...961..186C}, shown in cyan, green, blue, yellow, and red, respectively. The cyan box show the region of the mass distribution released by \citet{2024ApJ...961..186C}. Almost all the PAH emitters are in the region of mass surface density lower than 6 $\times 10^8\, M_\odot\,\rm kpc^{-2}$. 
    Our results show that PAH bright galaxies are more frequently seen in the cluster infall regions with lower surface density than its virialized core.
    }
    \label{A2744}
\end{figure*}

\subsection{Star formation rate of the PAH bright galaxies}

Since the PAH 3.3$\mu$m emitters are selected from the F430M photometry, we adopt our SExtractor dual mode photometry results to estimate the SFR$_{\rm PAH}$, which is optimal to the PAH emission region. We use the F444W AUTO flux as the dust continuum, and the flux$_{\rm F430M}$ - flux$_{\rm F444W}$ as PAH 3.3$\mu$m flux ($F_{\rm 3.3\mu m} [\rm erg\, s^{-1}\, cm^{-2}]$), omitting the other emission lines such as the Pfund $\delta$ line at 3.29 $\mu$m, the 3.4$\mu$m aliphatic feature, 3.4$\mu$m amorphous hydrocarbon (HAC) absorption \citep{2009ApJ...703..270S} and other emissions in the F430M filter. This will introduce an uncertainty about 10\% \citep{2025A&A...693A.204V}. The PAH flux of our sample is estimated as:
\begin{equation}\label{f430mexcess}
    F_{\rm 3.3\mu m} = \Delta {\rm F430M} \frac{f_{\rm F430M} - f_{\rm F444W} - f_{\lambda}^{\rm zpt}}{1-\Delta {\rm F430M}/\Delta {\rm F444W}},
\end{equation}
where $F_{\rm 3.3\mu m}$ is the line flux in units of $\rm erg\, s^{-1}\, cm^{-2}$, $f_{\rm F430M}$ and $f_{\rm F444W}$ are the flux densities in units of $\rm erg\, s^{-1}\, cm^{-2}\, \AA^{-1}$, and $f_{\lambda}^{\rm zpt} = 6.4\times10^8 \, \rm erg\, s^{-1}\, cm^{-2}\, \AA^{-1}$ represents the stellar continuum offset between F430M and F444W, estimated from the central value of the histogram of $f_{\rm F430M} - f_{\rm F444W}$ for cluster members with no F430M excess in A2744, obtained through Gaussian fitting. This value accounts for the intrinsic color of F444W - F430M caused by the SED slope \citep{2024arXiv241011808P}, and serves as the zeropoint for the flux excess in Equation \ref{f430mexcess}. The $\Delta \rm F430M = 2315.31\,\AA$, $\Delta \rm F444W = 11144.05\,\AA$ are the FWHMs of the F430M and F444W band response curves \citep{2011ApJ...726..109L, 2014ApJ...784..152A, 2018ApJ...864..145H}. 

We utilize the 3.3$\mu$m PAH and SFR correlation: $\log({\rm SFR_{\rm 3.3\mu mPAH}}/ M_\odot \, {\rm yr^{-1}}) = -(6.80 \pm 0.18) + \log(L_{\rm 3.3PAH}/L_\odot)-0.05$, calibrated by \citet{2020ApJ...905...55L, 2025A&A...693A.204V}. Since our targets are bright in NIRCam images, the flux uncertainties are much lower than the calibration error of the scaling relation, and therefore we adopt 0.18 dex as the 1-$\sigma$ uncertainty of SFR$_{\rm 3.3\mu mPAH}$.

We compare the SFR$_{\rm 3.3\mu mPAH}$ with the SFR from SED fitting results with Bagpipes and Magphys in Figure \ref{SFRPAH}. The SFR$_{\rm SED}$ from Bagpipes are mainly the SFR derived from the SEDs in rest-frame UV to near infrared, whereas the SFR$_{\rm SED}$ from Magphys also include the dust emission, and thus closer to the total SFRs. In Figure \ref{SFRPAH}, the SFR$_{\rm 3.3\mu mPAH}$ is similar to or higher than the SFR$_{\rm SED}^{\rm Bagpipes}$, while for the seven Herschel bright targets, SFR$_{\rm SED}^{\rm Magphys}$ is systematically higher than the SFR$_{\rm SED}^{\rm Bagpipes}$, and is consistent with SFR$_{\rm 3.3\mu mPAH}$ (blue circles in Figure \ref{SFRPAH}). Therefore, we conclude that the 3.3 $\mu$m PAH flux estimated from the medium-band photometry is well correlated with the SFR.

The SFRs measured from SED fitting and 3.3 $\mu$m PAH are shown in Figure \ref{ms}, as comparison to the results of the star-forming main sequence at $z = 0.3$ \citep{2014ApJS..214...15S}. The PAH emitters mainly have a similar or higher SFR as the field galaxies, while most of the targets in A2744 have low star formation rate. This shows that the PAH selection method will find more starburst galaxies even in clusters.

\begin{figure}[ht]
    \centering
    \includegraphics[width=0.95\linewidth]{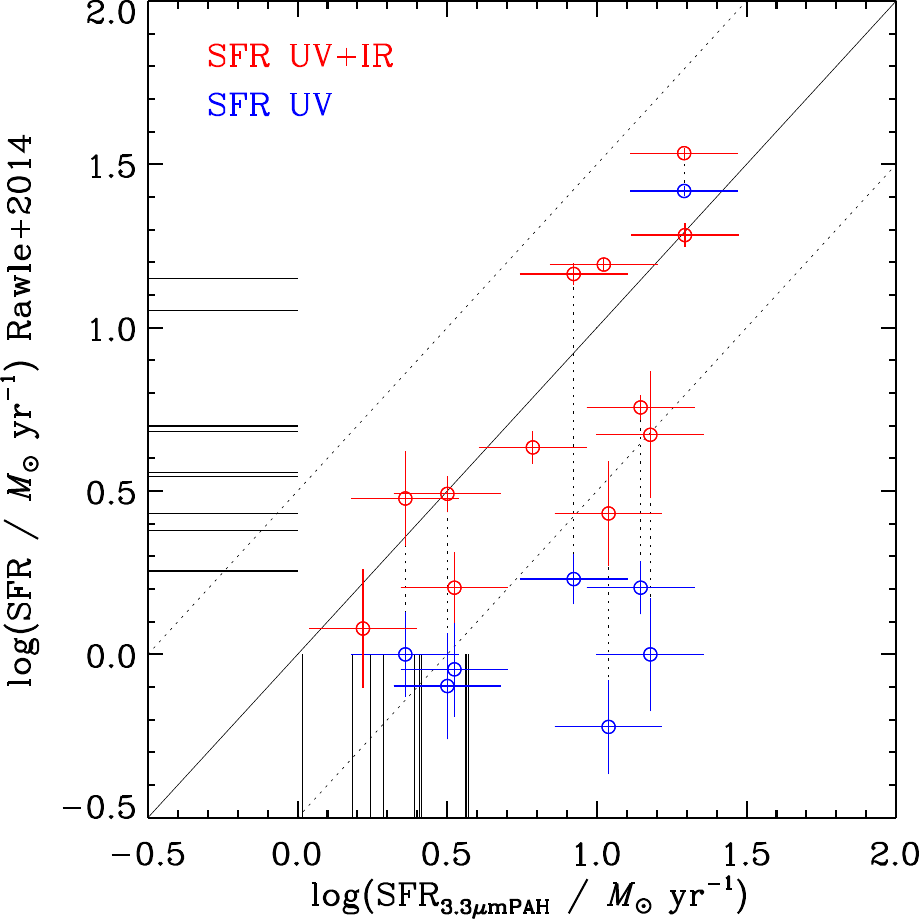}
    \caption{Comparison between the star forming galaxy sample in \citet{2014MNRAS.442..196R} and the 3.3 $\mu$m PAH sample in this work. The SFR of the y-axis are estimated from GALEX UV flux (blue) and UV+IR flux (red), where the IR flux are measured from MIPS or Herschel, which is roughly the total flux. As shown in Figure \ref{SFRPAH}, the SFR from UV+IR is consistent with the SFR$_{3.3 \mu \rm m PAH}$. We also denote the SFR of the targets that are only detected in \citet{2014MNRAS.442..196R} or in F430M by horizontal or vertical black bar in y or x-axis. The two star-forming galaxy samples are highly consistent with each other at SFR $> 10 M_\odot\rm\, yr^{-1}$. The two targets missed by our F430M method are GLX001421-302209 and HLS001414-302240, which are classified as further spiral and Red-core spiral in \citet{2014MNRAS.442..196R}. The SEDs of the two targets do not show excess in F430M.
    }
    \label{SFgalaxy}
\end{figure}

\subsection{PAH 3.3$\mu$m Morphology}

One advantage of our selection method is to reveal the PAH morphology (we assume the PAH morphology the same as F430M image subtract the F444W image). We compare the PAH morphology with the F444W morphology, which is the rest frame of 3.4 $\mu$m, and strongly correlated with the stellar morphology \citep[e.g., ][or Figure \ref{mass} in the Appendix]{2012AJ....143..139E}. We show the Gini-M20, Asymmetry index and half light radius in Figure \ref{morph}. The parameters are measured following the same method of \citet{2004AJ....128..163L}, with the segment maps generated by SExtractor.

Gini-M20 describes the morphology with non-parametrically method, and is widely used to quantify the flux concentration and possible tidal features, and classify the galaxy morphology \citep{2004AJ....128..163L, 2008ApJ...672..177L, 2012ApJ...752..134W, 2024ApJ...970...29L}. In the left panel of Figure \ref{morph}, we can see that the stellar distribution of our sample mainly has the merger feature, and the PAH distribution is closer to the feature of spiral galaxies, which suggests the PAH emission is more extended, and more disky for big galaxies.

Asymmetry of the stellar and PAH emission may hint at the gas ram-pressure in A2744 cluster. The asymmetric of PAH morphology is similar to the stellar morphology with a scatter about 0.1. However, it becomes noticeably more asymmetric when the asymmetry index exceeds 0.4 (Figure \ref{morph}, middle panel). We also compare the half-light radius of the PAH-bright region and that of the stellar component in the right panel of Figure \ref{morph}. The PAH-bright region appears more extended when $R_{\rm e, PAH} > 0.8''$ (approximately 3.6 kpc at $z = 0.308$), suggesting that larger galaxies tend to host more spatially extended star-forming regions.

Using the size measurements, we present the surface densities of stars ($\rm \Sigma_{\rm star} = \it M_{\rm star} \rm / 2\pi {\it R}e_{F444W}^2$) and star formation ($\rm \Sigma_{\rm SFR} = SFR / 2\pi {\it R}e_{PAH}^2$) in Figure \ref{SigmaSFR}, where ${\it R}e_{F444W}$, ${\it R}e_{PAH}$ are the half-light radius of F444W and PAH images. We compare the surface density with MAGPI survey project \citep{2024MNRAS.530.5072M}, which aims to study the star forming galaxies at $0.25<z<0.35$ with VLT/MUSE. The PAH sample predominantly lies above the scaling relation of MAGPI sample\footnote{
We note the $\Sigma_{\rm star}$ and $\Sigma_{\rm SFR}$ in \citet{2024MNRAS.530.5072M} are derived within the same diameter, while our definations are using the diameter of stellar and PAH morphology. From the right panel of Figure \ref{morph}, we can see the diameter difference between stellar and PAH images will not lead to the systematical offset in the $\Sigma_{\rm star} - \Sigma_{\rm PAH}$ distribution in Figure \ref{SigmaSFR}.}. A higher star formation surface density may indicate a higher HI surface density, and targets with high $\rm \Sigma_{\rm star}$ and $\rm \Sigma_{\rm SFR}$ are expected to have higher molecular gas surface densities \citep{2020MNRAS.496.4606M}, and higher metallicity \citep{2019MNRAS.484.5009E}. The surface density also helps to classify the star formation activity. In Figure \ref{SFRSigmaSFR}, we show the star formation surface density for normal/irregular galaxies, infrared-selected galaxies (such as ULIRGs), blue compact starburst galaxies, and circumnuclear star-forming rings in local barred galaxies from \citet{2012ARA&A..50..531K}. The 3.3 $\mu$m PAH targets are primarily between normal and infrared galaxies.

\subsection{Projected locations of the PAH bright galaxies}

We show the 3.3$\mu$m PAH selected target location in Figure \ref{A2744} with the the mass density contour from \citet{2024ApJ...961..186C}. Almost all the targets locate in the region with mass surface density lower than 6 $\times 10^8\, M_\odot\,\rm kpc^{-2}$, and clearly offset from the massive galaxies, consistent with the morphology density relation \citep{1980ApJ...236..351D, 1997ApJ...490..577D}. Gas temperature in the galaxy cluster center regions are high to about $10^7\, \rm K$, and would remove the cold gas from galaxies by ram pressure. Therefore, as the PAH bright galaxies infall toward the cluster central region, they are unlikely to acquire additional cold gas to sustain ongoing star formation. Consequently, the PAH sample may represent the most recent episode of star formation within the cluster environment.

We show the direction to the cluster center, and from the PAH image in Figure \ref{stamp1} and \ref{stamp2}. Most of the tail directions are not aligned with the cluster center. Since the star formation activity is usually in the high density region, the PAH morphology would not be quite sensitive to the ram pressure. Simulation also shows that the ram pressure tail does not always lie opposite to the galaxy cluster center \citep{2015MNRAS.449.2312V, 2017ApJ...841...38V, 2024MNRAS.533..341S}. 

Comparing the central direction and the asymmetry, we can see the target 5515 and 2763 have clear tails opposite to the cluster center, implying that these galaxies might just fall into the cluster from field. HI as the most diffuse baryonic components is the lightest elements and the most sensitive to the ram pressure. Highly sensitive telescopes such as MeerKAT are crucially important to show the ram pressure direction, and understand the interaction in the ICM \citep{2024MNRAS.533..341S}.

\subsection{Notes on Individual Galaxies}\label{target}

{\it ID 1911} has a clear point source in the center, and clumpy PAH morphology. This target is identified as a jellyfish galaxy and has been studied in detail with Chandra and optical spectrum by AAO \citep{2012ApJ...750L..23O}, Gemini/GMOS-IFU \citep{2022ApJ...940...24L} and spatial-resolved SED study with JWST/NIRCam images \citep{2024arXiv240915215W}. Since PAH emission would be destroyed by AGN, the PAH flux we estimated in this work may be closer to the SFR of the host galaxy.

{\it ID 2193} is a small galaxy with a neighbor galaxy at east (Figure \ref{stamp2}). This target is selected as F200W-F444W ``Red Excess'' galaxy in \citep{2023ApJ...948L..15V}. The JWST/NIRSpec spectrum of this target shows a clear PAH emission \citep[Figure 13 in][]{2023ApJ...948L..15V}, validating our selection method. One follow up study of this target with VLT/MUSE spectrum is in preparation (Hu et al. in prep).

{\it ID 1643, 2217, 5565, 6634, 2063} were also identified as F200W-F444W ``Red Excess'' galaxies in \citet{2023ApJ...948L..15V} or \citet{2025A&A...693A.204V}. So the excess may also be caused by the PAH emission as well as the existence of hot dust. 3.3 $\mu$m PAH emission lines of {\it ID 2217, 2217, 5565} are clearly shown in the NIRSpec spectra \citep{2025A&A...693A.204V}. Optical spectrum of {\it ID 2217} is shown in \citep{2012ApJ...750L..23O}, and identified as one jellyfish galaxy. The optical spectrum is classified as starburst \citep[Figure 2 lower right panel in ][]{2012ApJ...750L..23O}.

\section{Discussion} \label{sec:dis}

\subsection{Comparing with other SF galaxy sample in A2744}

\citet{2014MNRAS.442..196R} selected a sample of star forming galaxies from GALEX, Herschel and Spitzer/MIPS bright targets with spectroscopic redshift in A2744, and obtained a total star formation rate of of 201$\pm 9 M_\odot \rm yr^{-1}$ in the center 1.1 Mpc of A2744. The star formation rates in \citet{2014MNRAS.442..196R} are estimated from UV, IR or UV+IR when available. We cross match the our 3.3 $\mu$m PAH sample with the star forming sample in \citet{2014MNRAS.442..196R}, and show the results in Figure \ref{SFgalaxy}. For the 22 3.3 $\mu$m PAH bright galaxies in this work, we cross-matched 12 of them in both sample. As shown in Figure \ref{SFRPAH}, the SFR from F430M excess is consistent with the SFR estimated from UV+IR, indicating a highly completeness in SFR $> 10 M_\odot\rm\, yr^{-1}$. For the miss-matched targets in both sample, the SFR are mainly at $< 10 M_\odot\rm\, yr^{-1}$. The two targets GLX001421-302209 and HLS001414-302240 in \citet{2014MNRAS.442..196R} with SFR $> 10 M_\odot\rm\, yr^{-1}$ are missed in the F430M selected sample. The two galaxies are classified as spiral and Red-core spiral \citep[see the Figure 2 in ][]{2014MNRAS.442..196R}. The HST+JWST SEDs of the two targets do not show excess in F430M.

We also assess the completeness of our PAH-bright target selection in terms of far infrared. In the A2744 field, 38 targets have been detected by Herschel \citep{2016MNRAS.459.1626R}. Cross-matching these sources, we find that all Herschel-bright galaxies at a photometric redshift of approximately 0.3 are also 3.3$\mu$m PAH emitters. This confirms that our 3.3$\mu$m PAH selection method is highly complete in detecting massive dusty galaxies and is even more sensitive to fainter dusty galaxies (Figure \ref{SFRPAH}). Moreover, as a method to probe the dusty galaxies, another advantage of the PAH selection method is its ability to reveal dust morphology (Figures \ref{stamp1} and \ref{stamp2}).

Our results indicate that the star formation rate estimated from optical to far-infrared SEDs is consistent with the SFR derived from 3.3$\mu$m PAH emission. This suggests a strong connection between the hot and cold dust components. High-resolution interferometric observations in the FIR-to-submillimeter continuum could further elucidate this relationship. Since the FIR-based SFR is sensitive to star formation within the past 100 Myr \citep{2012ARA&A..50..531K, 2012AJ....144....3L}, the morphological similarity between PAH emission and the FIR continuum could provide constraints on the timescale of the SFR$_{\rm 3.3\mu mPAH}$.

\begin{figure}[ht]
    \centering
    \includegraphics[width=0.96\linewidth]{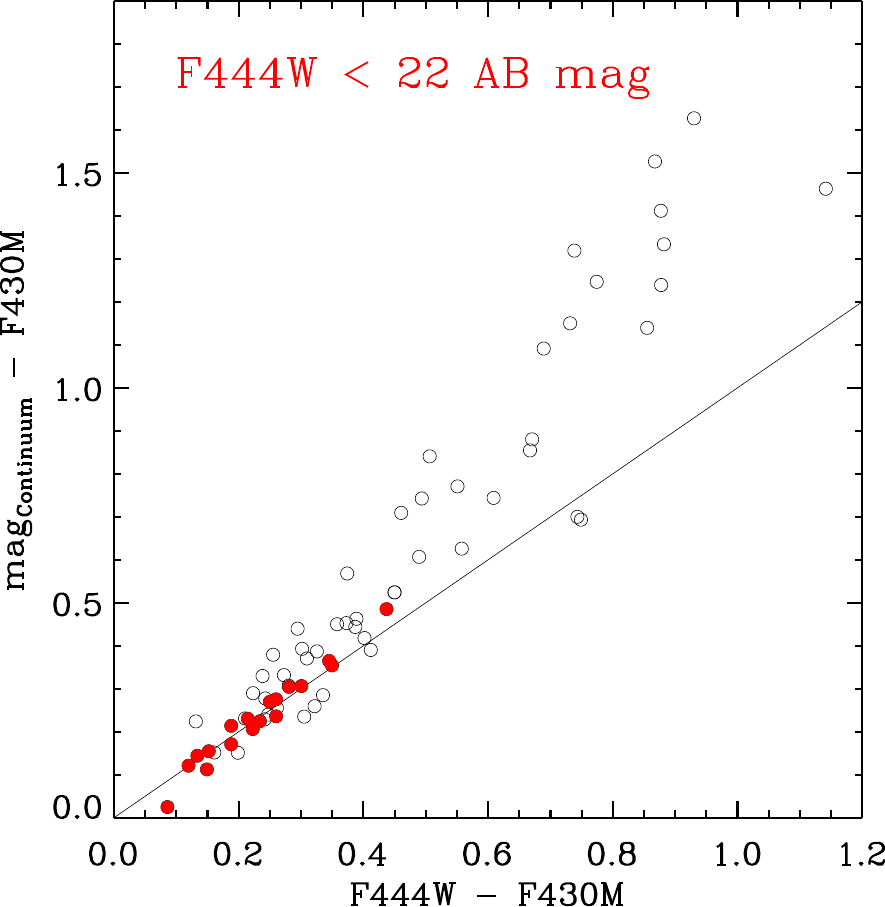}
    \caption{${\rm mag}_{\rm continuum} - \rm F430M$ vs. F444W - F430M for the F430M emitters, where ${\rm mag}_{\rm continuum} = 28.9 - 2.5 \times \log(0.6 \times f_{\rm F410M} + 0.4 \times f_{\rm F460M})$. Since the F444W flux includes the emission lines captured by F430M, and ${\rm mag}_{\rm continuum}$ more closely represents the line-free continuum for the F430M emitters, the color F444W - F430M is offset from ${\rm mag}_{\rm continuum} - \rm F430M$ for more extreme emitters when F444W - F430M $\gtrsim 0.4$. Therefore, assuming F444W as the continuum of the emission line lowers the significance of emitter selection. We highlight the PAH emitters in this work with red dots, which are closer to the 1:1 line in black. The scatter of the color difference for the PAH emitters is 0.022, which has little effect on PAH emitter selection. The scatter in color difference introduces an uncertainty in PAH flux of approximately $\Delta \log (f_{\rm PAH}) \sim 0.02 - 0.04$, depending on the F444W - F430M color (see Section \ref{bias}).
    }
    \label{f444wcont}
\end{figure}

\begin{figure}[ht]
    \centering
    \includegraphics[width=0.96\linewidth]{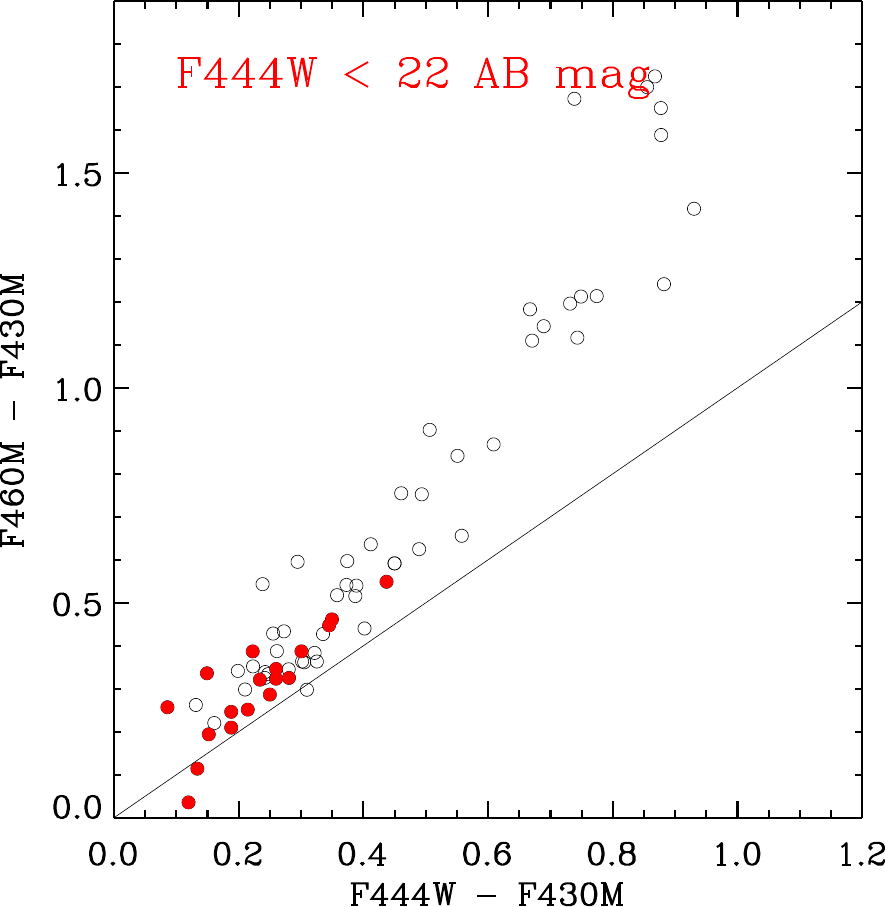}
    \caption{F460M - F430M vs. F444W - F430M for the F430M emitters. The solid line show the 1:1 trace. We highlight the 3.3$\mu$m PAH emitter by red dots as in Figure \ref{f444wcont}. The offset is caused by the intrinsic color between F444W and F460M, which suggests the continuum for the 3.3$\mu$m PAH emission should cover the wavelength blue and reder than F430M. The scatter of the color difference between F460M - F430M and F444W - F430M is 0.069, which will affect the accuracy of the PAH flux by 0.1 dex, and still not affect the selection and SFR estimation. 
    }
    \label{f460mcont}
\end{figure}

\subsection{Bias of the PAH selection method}\label{bias}

\subsubsection{Using F444W as the continuum of the 3.3 $\mu$m PAH emission}

We make use of F444W as the continuum for the 3.3 $\mu$m PAH emission to achieve a wider area of coverage in the image (Figure \ref{filter}). However, the emission line within F444W contaminates the continuum, especially when the equivalent width of the emission line is high. To address this issue, we estimate the continuum flux by interpolating the flux from F410M and F460M in the survey area where both bands are covered. We define the continuum magnitude as ${\rm mag}_{\rm continuum} = 28.9 - 2.5 \times \log(0.6 \times f_{\rm F410M} + 0.4 \times f_{\rm F460M})$, and compare ${\rm mag}_{\rm continuum} - \rm F430M$ with F444W - F430M in Figure \ref{f444wcont}.
The weights (0.6 and 0.4) are based on the relative distances of the F430M central wavelength (4.3 $\mu$m) to the neighboring bands F410M (4.1 $\mu$m) and F460M (4.6 $\mu$m).
When the F444W - F430M color excess is higher than about 0.4, there is a clear offset between ${\rm mag}_{\rm continuum} - \rm F430M$ and F444W - F430M, indicating that the flux difference between F430M and F444W biases the intrinsic emission line flux. Therefore, targets with a high F444W - F430M color excess will have a lower significance in target selection.

We highlight PAH emitters with coverage in both the F410M and F460M data in Figure \ref{f444wcont} and find that all PAH emitters in this work have colors close to the 1:1 line, with a color difference scatter of 0.022. The F444W - F430M color is defined as $2.5\log(f_{\rm F430M}/f_{\rm F444W})$. Assuming $ f_{\rm PAH} = f_{\rm F430M} - f_{\rm F444W} = 10^{\rm color/2.5} \times f_{\rm F444W} - f_{\rm F444W}$, the color difference introduces an uncertainty of approximately $\Delta f_{\rm PAH} = (1/2.5) \times \ln(10) \times 0.022 \times f_{\rm F444W} = 0.02f_{\rm F444W}$, corresponding to $\Delta\log(f_{\rm PAH}) = \Delta f_{\rm PAH} / f_{\rm {PAH}} / \ln(10) = 0.02 / (10^{{\rm color}/2.5} - 1) / \ln(10)$. This results in a value of about 0.02 to 0.04 for F444W - F430M colors ranging from 0.4 to 0.2, which is significantly smaller than the typical scatter in the scaling relation between $f_{\rm PAH}$ and SFR. Therefore, our approximation of using F444W flux as the continuum for selecting PAH emission does not significantly affect the accuracy of selection significance or SFR values. This is partly because the PAH targets in this work are massive and bright in F444W flux, and thus do not exhibit extremely high equivalent widths for the 3.3 $\mu$m PAH emission.

Meanwhile, since F410M includes part of the 3.3$\mu$m PAH emission, while F460M serves as a more {\it line-free} filter for the PAH feature (Figure~\ref{filter}), we further assess the uncertainty of using F444W as the continuum estimate in Figure~\ref{f460mcont} by adopting F460M as a continuum. The offset between F444W$-$F430M and F460M$-$F430M arises from the intrinsic color of F444W$-$F460M, highlighting the importance of estimating the continuum from both the blue and red sides of F430M. The scatter between F444W$-$F430M and F460M$-$F430M is 0.069 mag, corresponding to $\Delta\log(f_{\rm PAH}) \sim 0.14 - 0.06$ for the F444W$-$F430M of 0.2 to 0.4, which does not significantly affect the target selection and SFR results in this work.

\subsubsection{The bias of 3.3 $\mu$m PAH emitters to the low mass star forming galaxies}

PAH emission in dwarf galaxies and low-metallicity environments is known to be faint, as reported in several studies \citep{2004ApJS..154..211H, 2005ApJ...624..162H, 2005ApJ...628L..29E, 2006ApJ...639..157W, 2011ApJ...730..111W, 2024ApJ...974...20W, 2024A&A...690A..89S}. This faintness may be due to the low abundance of carbon in such environments, leading to the formation of smaller PAH molecules that are more susceptible to destruction \citep[e.g.,][]{2010ApJ...715..701S, 2012ApJ...744...20S}. Meanwhile, PAHs that can be more easily destroyed by strong UV radiation fields in dwarf galaxies where there were few dust to shield the strong radiation \citep{2003A&A...407..159G, 2005A&A...434..867G, 2006A&A...446..877M}. Especially for the 3.3$\mu$m feature, which arises from smaller neutral PAHs than other PAH features. Alternatively, some dwarf galaxies may still be too young to have formed PAHs (from Figure \ref{SFRSigmaSFR}, we can see that the blue compact galaxies and our PAH sample are separated). As a result, selecting star-forming galaxies based on PAH emission or medium-band photometry tends to bias the sample toward more massive galaxies. In this study, we focus on massive galaxies with recent star formation and apply a flux limit of F444W $<$ 22 AB mag, thereby avoiding the issue of PAH deficiency in dwarf galaxies.

How can we select a more complete star-forming galaxy sample in one galaxy cluster? Given the PAH deficit in dwarf galaxies, the SED or UV selection method may be more effective in capturing the low-SFR galaxy population more completely. From Figure \ref{SFRPAH}, we can see that when the SFR is approximately $\sim 1 M_\odot \,\rm yr^{-1}$, the PAH and SED methods could yield similar SFR estimates. The low dust abundance in dwarf galaxies results in low dust extinction correction and leads to a more reliable SFR from SED fitting or direct measurement from UV flux. If spectroscopic redshift data is available, a galaxy sample selected using the HST UV image combined with the medium band at 3.3 $\mu$m PAH will be highly complete in terms of star formation rate estimation, as well as providing insight into the spatial distribution of star formation.

\subsection{Projected location of the Star-forming galaxies in Cluster: Connection to the Cosmic Filaments?}

Observations of massive galaxy clusters also show the filamentary structures \citep{2016ApJ...833..207K, 2021ApJ...906...68L, 2021ApJ...923..235C}, indicating the large scale structures of the universe, as well as the gas accretion into massive halos. Previous X-ray observations of A2744 have shown three main filaments in the east, north-west and south direction \citep{2007A&A...470..425B, 2015Natur.528..105E, 2024A&A...692A.200G}. These filaments connect to field galaxies, and thus the filament direction would have more star forming galaxies. 
The 3.3 $\mu$m-bright galaxies in A2744 might be more closely connected to the nearby field galaxies and are possibly infalling into the cluster along filaments. We can also expect the position of the star forming galaxies in clusters would be the end point of the cosmic filaments toward galaxy clusters \citep{2024A&A...692A.200G, 2024arXiv241113655S}. 

Our target selection method can identify star-forming galaxies efficiently. However, Limited by the F430M coverage area, our PAH sample cannot trace more wide area of the galaxies at $z = 0.308$. Spectroscopic redshift survey project such as DESI would detect more targets at $z \simeq 0.3$, and would provide a clearer view of the filamentary structures around A2744.

\subsection{Why Are They Still Star-forming?}

The cold gas in star-forming galaxies is very likely to be ram-pressured or heated by the intercluster medium, and as a result, the galaxies are quenched. Our SED fitting results for the cluster members indicate that the main stellar population is formed at 4 Gyr after the Big Bang. This may also be the formation time of the core region of the galaxy cluster, while the follow-up mergers keep building up A2744 \citep{2011MNRAS.417..333M, 2011ApJ...728...27O}.

The quenching of recent star-forming galaxies in galaxy clusters may be related to the timescale of their entry into the cluster, or the number of interactions they have experienced. As shown in Figure \ref{SFH}, the non-parametric SFH of our sample exhibits a recent starburst peak around $z \simeq 0.3$ or 8 to 10 Gyr, which may indicate recent star formation followed by rapid quenching. Cold gas in clusters can be exhausted by ICM heating, ram pressure, or tidal stripping from galaxies, and may not support continuous star formation since the formation of the cluster. This is consistent with the lack of PAH emitters in Figure \ref{A2744} in high mass surface density region. Thus, the PAH sample would have a long gas depletion timescale, or more likely, these galaxies have only recently entered the cluster from field. Then their star formation activity may remain largely unchanged until they are eventually quenched (Figure \ref{ms}). On the other hand, the lack of star forming galaxy at high mass density region also suggest a quick quenching process when field galaxies fall in the cluster. Comparison between the mass surface density and the distribution of the Post-starburst galaxies identified from MUSE or ATT spectra will help to constrain the effect of quenching in ICM.

Moreover, to understand the star formation properties of the PAH selected sample, we still need to estimate the gas consumption timescale, and thus the low-J CO observations of this PAH-bright sample are crucial for understanding the quenching process in clusters. 

\section{Summary} \label{sec:sum}

We present a sample of 3.3$\mu$m PAH-bright galaxies in the A2744 galaxy cluster. Using F430M medium band images, we select PAH emitters at the redshift of A2744. We find that the star formation rates derived from both the 3.3$\mu$m PAH flux estimated from medium band image and UV-to-FIR SED fitting are consistent, demonstrating that our PAH selection from medium band images is efficient and reliable, particularly for identifying dusty star-forming galaxy population. The star formation rate of our sample aligns with the star-forming galaxy main sequence, suggesting that the star formation activity in galaxy clusters is similar to that of field galaxies. 

One advantage of the PAH selection method is to reveal the dust-free star formation rate and star formation size simultaneously. We find that the size of the PAH emission region is either similar to or larger than the F444W image, suggesting a more extended star formation mode for larger galaxies, similar to that of spiral galaxies.

The non-parametric SFH results of the PAH emitters show a recent starburst peak. Meanwhile, the PAH emitters are primarily located in the low mass density region ($<6 \times 10^8\, M_\odot\,\rm kpc^{-2}$) of A2744. The consistency with the star-forming main sequence, the absence of PAH emitters in high mass density regions, the recent starburst indicated by the non-parametric SFH, and the asymmetry in the PAH morphology suggest that the PAH-selected star-forming galaxies in clusters have recently fallen into the cluster from the field.

Previous studies have highlighted the filamentary structures around A2744. The star-forming galaxies in A2744 may reside at the endpoints of cosmic filaments feeding into the cluster, with star formation activity potentially influenced by the surrounding intracluster medium. In addition to identifying star-forming galaxies in A2744, we suggest that our findings could point to windows towards these filaments.

The medium band imaging from JWST offers a new opportunity to identify emission-line galaxies and explore star-forming galaxies within galaxy clusters (e.g., F460M in M0416). Follow-up high-resolution HI and low-J CO observations will further enhance our understanding of the ram pressure and star formation activity in galaxy clusters.

\begin{longrotatetable}
\begin{deluxetable*}{lcccccccccc}
\tabletypesize{\scriptsize}
\tablenum{1}
\tablecaption{3.3$\mu$m PAH sample in A2744\label{tab}}
\tablehead{
\colhead{ID} & \colhead{RA} & \colhead{Dec} & \colhead{$z_{\rm spec}$} & \colhead{log(Mstar)} & F$_{3.3\rm \mu m PAH}$ & \colhead{$\log({\rm SFR}_{3.3\mu\rm m PAH})$} & \colhead{$\log({\rm SFR}_{\rm SED}^{\rm Bagpipes})$}  & \colhead{$\log({\rm SFR}_{\rm SED}^{\rm Magphys})$} & \colhead{UNCOVER ID\tablenotemark{d}} & \colhead{Herschel ID\tablenotemark{c}} \\
\colhead{} & \colhead{J2000} & \colhead{J2000} & \colhead{} & \colhead{$\log(M_\odot)$} & \colhead{$\rm \times 10^{-16}\,erg\,s^{-1}\,cm^{-2}$} & \colhead{$\log(M_\odot \, \rm yr^{-1})$} & \colhead{$\log(M_\odot \, \rm yr^{-1})$}  & \colhead{$\log(M_\odot \, \rm yr^{-1})$} & \colhead{} & \colhead{} \\
}
\decimalcolnumbers
\startdata
1191 & 00:14:26.6 & -30:23:44.2  &   0.3030\tablenotemark{a} &    9.14 $\pm$ 0.03   &   15.37 &     1.29   $\pm$  0.18 &   1.215 $\pm$ 0.001   &     1.457 $\pm$ 0.013 &  160916 &  HLSJ001426.6--302344 \\
2063 & 00:14:22.4 & -30:23:03.7  &   0.2962\tablenotemark{a} &   10.26 $\pm$ 0.03   &   15.46 &     1.28   $\pm$  0.18 &   0.669 $\pm$ 0.004   &     1.087 $\pm$ 0.085 &   23405 & HLSJ001422.4--302304 \\
3388 & 00:14:21.0 & -30:22:16.6  &   0.3040\tablenotemark{a} &   10.41 $\pm$ 0.03   &   8.28  &     1.02   $\pm$  0.18 &   0.473 $\pm$ 0.005   &     0.902 $\pm$ 0.103 &   33854 & HLSJ001421.0--302216 \\
5661 & 00:14:23.1 & -30:20:53.7  &   0.2887\tablenotemark{a} &   10.10 $\pm$ 0.03   &   10.99 &     1.15   $\pm$  0.18 &   0.030 $\pm$ 0.013   &      --               &  44796 &   -- \\
1643 & 00:14:19.4 & -30:23:26.8  &   0.2926\tablenotemark{a} &    9.98 $\pm$ 0.03   &   6.57  &     0.92   $\pm$  0.18 &   0.532 $\pm$ 0.003   &     1.067 $\pm$ 0.103 &   19562 & HLSJ001419.4--302327 \\
3059 & 00:14:03.6 & -30:22:24.4  &   0.3064\tablenotemark{b} &    9.96 $\pm$ 0.03   &   2.86  &     0.56   $\pm$  0.18 &  -0.112 $\pm$ 0.007   &      --               &  32278 &   -- \\
1536 & 00:14:28.5 & -30:23:34.5  &   0.3020\tablenotemark{a} &    9.01 $\pm$ 0.03   &   4.80  &     0.79   $\pm$  0.18 &   0.293 $\pm$ 0.002   &     0.682 $\pm$ 0.075 &   19205 & HLSJ001428.5--302334 \\
2193 & 00:14:25.1 & -30:23:05.8  &   0.2960\tablenotemark{a} &    9.23 $\pm$ 0.03   &   1.38  &     0.24   $\pm$  0.18 &  -0.136 $\pm$ 0.004   &      --               &  22353 &   -- \\
2217 & 00:14:16.6 & -30:23:03.2  &   0.2960\tablenotemark{a} &    9.25 $\pm$ 0.03   &   2.49  &     0.50   $\pm$  0.18 &   0.137 $\pm$ 0.002   &     0.347 $\pm$ 0.095 &   22890 & HLSJ001416.7--302304 \\
2432 & 00:14:07.3 & -30:22:50.2  &     --                    &    9.79 $\pm$ 0.03   &   1.30  &     0.22   $\pm$  0.18 &  -0.262 $\pm$ 0.011   &      --               &  25582 &   -- \\
2698 & 00:14:26.3 & -30:22:43.6  &   0.3007\tablenotemark{b} &    8.83 $\pm$ 0.03   &   2.00  &     0.41   $\pm$  0.18 &  -0.010 $\pm$ 0.003   &      --               &  26115 &   -- \\
5515 & 00:14:08.9 & -30:21:06.4  &     --                    &    9.66 $\pm$ 0.03   &   2.04  &     0.41   $\pm$  0.18 &   0.079 $\pm$ 0.007   &      --               &  42680 &   -- \\
6682 & 00:14:17.7 & -30:19:48.4  &     --                    &    9.86 $\pm$ 0.03   &   1.52  &     0.29   $\pm$  0.18 &  -0.301 $\pm$ 0.015   &      --               &  52391 &   -- \\
0621 & 00:14:25.4 & -30:24:35.0  &     --                    &    9.01 $\pm$ 0.03   &   0.18  &    -0.64   $\pm$  0.18 &  -0.529 $\pm$ 0.005   &      --               &  10121 &   -- \\
2763 & 00:14:21.3 & -30:22:37.2  &  0.2955\tablenotemark{c}  &    9.47 $\pm$ 0.03   &   1.20  &     0.18   $\pm$  0.18 &  -0.520 $\pm$ 0.004   &      --               &  27693 &   -- \\
2896 & 00:14:23.3 & -30:22:36.2  &     --                    &    9.24 $\pm$ 0.03   &   0.81  &     0.02   $\pm$  0.18 &  -0.023 $\pm$ 0.002   &      --               &  27705 &   -- \\
5565 & 00:13:53.3 & -30:21:01.1  &  0.3068\tablenotemark{b}  &   10.18 $\pm$ 0.03   &   2.92  &     0.57   $\pm$  0.18 &  -0.091 $\pm$ 0.009   &      --               &  43756 &   -- \\
6634 & 00:13:55.7 & -30:19:54.8  &     --                    &    9.84 $\pm$ 0.03   &   1.93  &     0.39   $\pm$  0.18 &  -0.116 $\pm$ 0.014   &      --               &  51697 &   -- \\
3599 & 00:13:53.7 & -30:22:12.2  &  0.3129\tablenotemark{b}  &    9.52 $\pm$ 0.03   &   1.80  &     0.36   $\pm$  0.18 &  -0.093 $\pm$ 0.004   &      --               &  163712 &   -- \\
1989 & 00:13:48.3 & -30:23:01.6  &  0.2910\tablenotemark{a}  &   10.05 $\pm$ 0.03   &   11.85 &     1.19   $\pm$  0.18 &   0.113 $\pm$ 0.006   &     0.642 $\pm$ 0.150 &  161941 &  HLSJ001348.0--302304 \\
1761 & 00:13:51.1 & -30:23:21.2  &  0.2906\tablenotemark{b}  &    9.86 $\pm$ 0.03   &   2.62  &     0.52   $\pm$  0.18 &  -0.196 $\pm$ 0.006   &      --               &  161470 &   -- \\
0737 & 00:13:49.7 & -30:24:19.8  &  0.2853\tablenotemark{b}  &   10.58 $\pm$ 0.03   &   8.59  &     1.04   $\pm$  0.18 &  -0.337 $\pm$ 0.008   &      --               &  160272 &   -- \\
\enddata
\tablenotetext{a}{$z_{\rm spec}$ from \citet{2022ApJS..263...38K}, which collects a wide range of spectroscopic surveys.}
\tablenotetext{b}{
$z_{\rm spec}$ from \citet{2011ApJ...728...27O}.The spectroscopic redshift of ID 2698 is 0.2389 in \citet{2022ApJS..263...38K} while 0.3007 in \citet{2011ApJ...728...27O}. We adopt the $z_{\rm spec} = 0.3007$ from \citet{2022ApJS..263...38K}, which is consistent with the F430M flux excess.
}
\tablenotetext{c}{
$z_{\rm spec}$ from \citet{Foex2017}.
}
\tablenotetext{d}{UNCOVER IDs from \citet{2024ApJS..270....7W}.}
\tablenotetext{e}{Herschel IDs are adopted from \citet{2016MNRAS.459.1626R}.}
\tablecomments{For the targets with no spectroscopic redshifts, we take their photometric redshift from UNCOVER in the SED fitting.}
\end{deluxetable*}
\end{longrotatetable}

\begin{acknowledgments}

We thank the anonymous referee for helpful and constructive comments that improved the clarity and quality of this paper. We are also grateful to Aigen Li, Benedetta Vulcani, Karl Glazebrook, Fuyan Bian, Zhiyu Zhang for insightful discussions and valuable suggestions during the development of this work.

All the HST and JWST data used in this paper can be found in MAST: \dataset[https://doi:10.17909/1esp-hh29]{https://doi:10.17909/1esp-hh29}. We would like to thank the MAGNIF project for providing valuable insights that inspired part of this work.

This work is sponsored (in part) by the Chinese Academy of Sciences (CAS) through a grant to the CAS South America Center for Astronomy. C.C. acknowledges NSFC grant No. 11803044 and 12173045. This work is supported by the China Manned Space Program with grant no. CMS-CSST-2025-A07. C.C. is supported by Chinese Academy of Sciences South America Center for Astronomy (CASSACA) Key Research Project E52H540301.

X. W. is supported by the National Natural Science Foundation of China (grant 12373009), the CAS Project for Young Scientists in Basic Research Grant No. YSBR-062, the Fundamental Research Funds for the Central Universities, the Xiaomi Young Talents Program, and the China Manned Space Program with grant no. CMS-CSST-2025-A06. X. W. also acknowledges work carried out, in part, at the Swinburne University of Technology, sponsored by the ACAMAR visiting fellowship.

E.I. gratefully acknowledge financial support from ANID - MILENIO - NCN2024\_112 and ANID FONDECYT Regular 1221846.
J.M. gratefully acknowledge financial support from ANID - MILENIO - NCN2024\_112.

(Some of) The data products presented herein were retrieved from the Dawn JWST Archive (DJA). DJA is an initiative of the Cosmic Dawn Center (DAWN), which is funded by the Danish National Research Foundation under grant DNRF140.

\end{acknowledgments}

\begin{contribution}

The project was conceived by Cheng Cheng and Xin Wang. Cheng Cheng performed the data analysis and wrote the manuscript. Piaoran Liang carried out the morphological analysis. All authors contributed to the discussion and interpretation of the results.


\end{contribution}

%

\facilities{HST, JWST, {\it Herschel}, ALMA}


\software{astropy \citep{2013A&A...558A..33A,2018AJ....156..123A},
          Source Extractor \citep{1996A&AS..117..393B}
          }


\appendix

\section{Observed magnitude and the stellar mass}

Rest frame near infrared flux are mainly from low mass stars, and thus correlated well with the galaxy stellar mass. We show the stellar mass and the observed Ks band and IRAC/ch2 band magnitude for the galaxies selected from CANDELS catalog with $0.25<z_{\rm spec}<0.35$. The tight correlation indicate that galaxies with F444W$<22$ or F200W$<20$ AB mag would be galaxies with $M_{\rm star}/M_\odot > 10^9$. The stellar mass of our sample from SED fitting is also consistent with the mass-light relation shown in Figure \ref{mass}.

\setcounter{figure}{0}
\renewcommand{\thefigure}{A\arabic{figure}}

\begin{figure}
    \centering
    \includegraphics[width=0.5\linewidth]{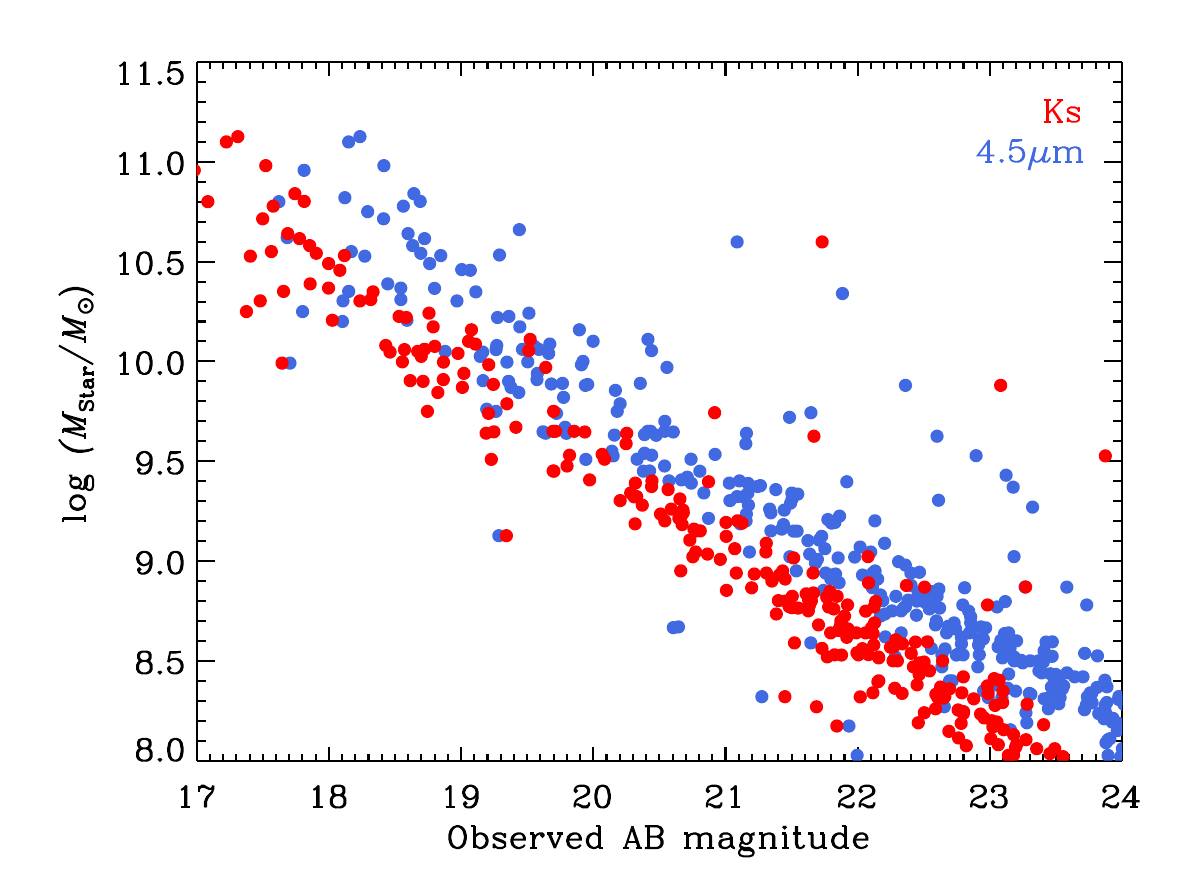}
    \caption{Stellar mass and the observed Ks and 4.5$\mu$m bands for the galaxies at $0.2<z<0.4$ from CANDELS catalog. The strong correlation indicate that the flux cut in near infrared flux limited sample is also a stellar mass limited sample.}
    \label{mass}
\end{figure}

\section{SFR from different SFH assumption}
The SFR from SED fitting is highly depended on the assumption of SFH. To verify the SFR$_{\rm SED}$, we compare the SFR$_{\rm SED}$ from the star formation history of double power law and non-parameter in Figure \ref{SFR_SFH}. 

\section{SFR from Herschel data and MAGPHYS fitting results}

For the six targets detected by Herschel, we compare the SFR derived from Herschel data by \citet{2016MNRAS.459.1626R} and the SFR from MAGPHYS fitting in Figure \ref{sfr_herschel}. The results are consistent except for ID 1191 with a much higher $SFR_{\rm MAGPHYS}$ value, which is caused by the AGN contamination in optical blue bands of SED.

\setcounter{figure}{0}
\renewcommand{\thefigure}{B\arabic{figure}}

\begin{figure}
    \centering
    \includegraphics[width=0.5\linewidth]{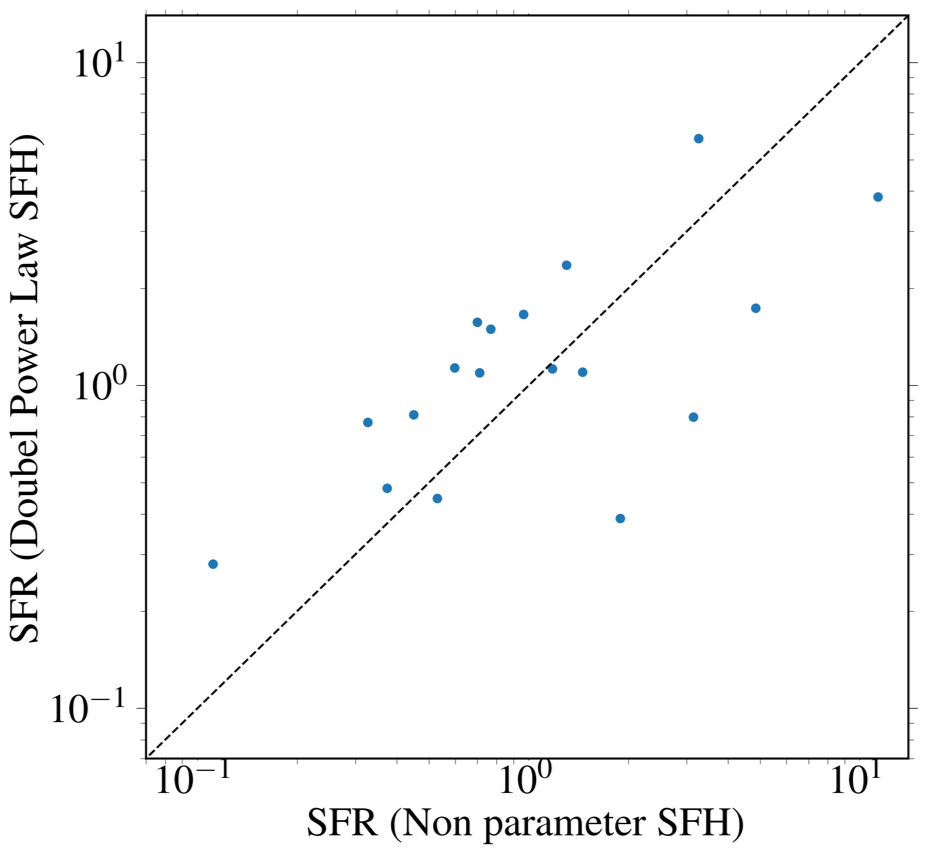}
    \caption{The SFR from the SED fitting would be affected by the uncertainty of SFH. We compare the SFR from Bagpipes with double power law SFH and the non-parameter SFH, and find a consistent trend of the two SFR$_{\rm SED}$.
    }
    \label{SFR_SFH}
\end{figure}

\setcounter{figure}{0}
\renewcommand{\thefigure}{C\arabic{figure}}

\begin{figure}
    \centering
    \includegraphics[width=0.5\linewidth]{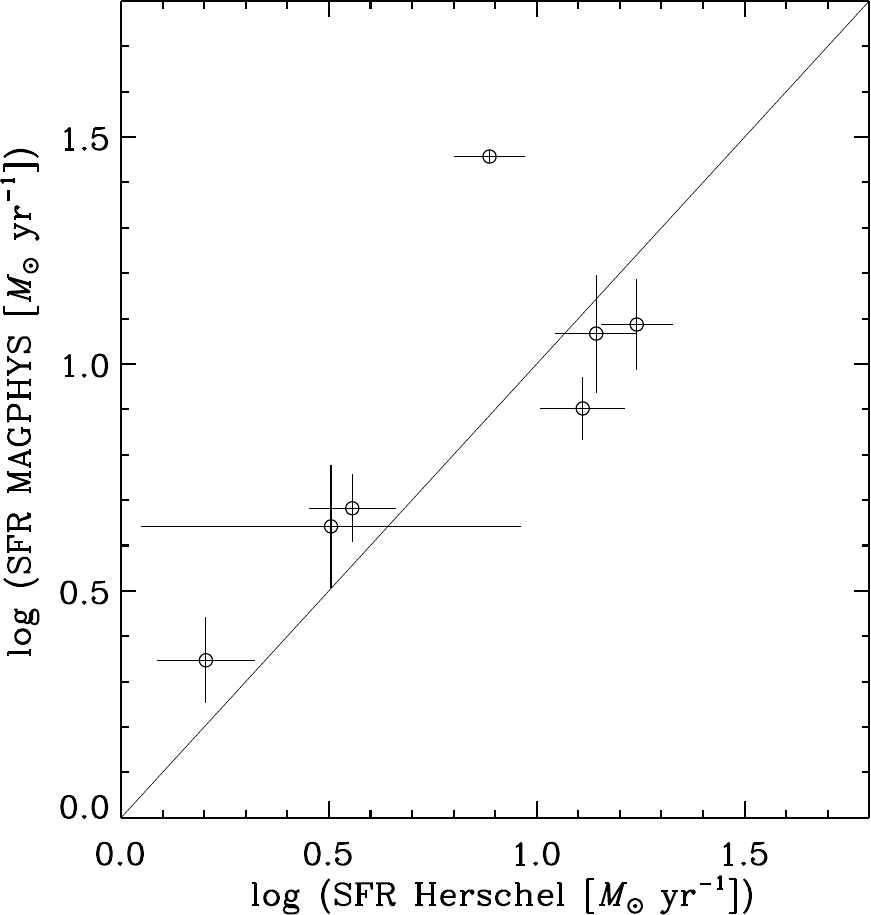}
    \caption{Comparison between the SFR derived from Herschel data \citep[SFR$_{\rm Herschel}$ adopted from][]{2016MNRAS.459.1626R} and from the SED fitting (SFR$_{\rm MAGPHYS}$). The target with the highest SFR$_{\rm MAGPHYS}$ is ID 1191, which includes an AGN in the center, and bias the SFR from SED fitting.
    }
    \label{sfr_herschel}
\end{figure}





\end{document}